\documentclass[letterpaper]{article} 
\usepackage{aaai24}  
\usepackage{times}  
\usepackage{helvet}  
\usepackage{courier}  
\usepackage[hyphens]{url}  
\usepackage{multirow}
\usepackage{float}
\usepackage{svg}
\usepackage{subcaption}
\usepackage{graphicx} 
\usepackage{natbib}  
\usepackage{caption} 
\usepackage{amssymb}
\usepackage{dsfont}
\usepackage{amsmath}
\usepackage{algorithm}
\usepackage{algorithmic}
\newtheorem{definition}{Definition}
\usepackage{newfloat}
\usepackage{listings}
\DeclareCaptionStyle{ruled}{labelfont=normalfont,labelsep=colon,strut=off} 
\floatstyle{ruled}
\newfloat{listing}{tb}{lst}{}
\floatname{listing}{Listing}
\title{No prejudice! Fair Federated Graph Neural Networks for Personalized Recommendation}
\author {
    Nimesh Agrawal\textsuperscript{\rm 1}\equalcontrib,
    Anuj Kumar Sirohi\textsuperscript{\rm 2}\equalcontrib,
    Sandeep Kumar\textsuperscript{\rm 1,2},
    Jayadeva\textsuperscript{\rm 1,2}
}
\affiliations {
     \textsuperscript{\rm 1} Department of Electrical Engineering, Indian Institute of Technology, Delhi, India\\
     \textsuperscript{\rm 2} Yardi School of Artificial Intellgence, Indian Institute of Technology, Delhi, India\\
    nimeshagrawal84@gmail.com, aiz218324@iitd.ac.in,
     ksandeep@ee.iitd.ac.in, jayadeva@ee.iitd.ac.in,
   
}

\begin{document}

\maketitle

\begin{abstract}
Ensuring fairness in Recommendation Systems (RSs) across demographic groups is critical due to the increased integration of RSs in applications such as personalized healthcare, finance, and e-commerce. Graph-based RSs play a crucial role in capturing intricate higher-order interactions among entities. However, integrating these graph models into the Federated Learning (FL) paradigm with fairness constraints poses formidable challenges as this requires access to the entire interaction graph and sensitive user information (such as gender, age, etc.) at the central server. This paper addresses the pervasive issue of inherent bias within RSs for different demographic groups without compromising the privacy of sensitive user attributes in FL environment with the graph-based model. To address the group bias, we propose F$^{2}$PGNN (\textbf{F}air \textbf{F}ederated \textbf{P}ersonalized \textbf{G}raph \textbf{N}eural \textbf{N}etwork), a novel framework that leverages the power of Personalized Graph Neural Network (GNN) coupled with fairness considerations. Additionally, we use differential privacy techniques to fortify privacy protection. Experimental evaluation on three publicly available datasets showcases the efficacy of F$^{2}$PGNN in mitigating group unfairness by $47\% \sim 99\%$ compared to the state-of-the-art while preserving privacy and maintaining the utility. The results validate the significance of our framework in achieving equitable and personalized recommendations using GNN within the FL landscape.  
\end{abstract}

\section{Introduction} \label{sec:Introduction}
Online recommendation systems (RSs) are used in various platforms in the modern market, such as e-commerce, e-learning, music and movie recommendation to targeted individuals/audiences \cite{sarwar2000}. Traditional RSs collect user data on a centralized server, which entails serious privacy and security issues. Machine learning (ML) models can now be locally trained thanks to edge devices' growing storage and processing capabilities. Due to this, Federated Learning (FL) has emerged, allowing clients to communicate updates with the server without actually sending any data \cite{mcmahan2017}. The server then suggests a global model, which is shared with every client. Clients train locally using their data and transmit the updated model back to the server for aggregation. In recent years, FL has been successfully used in various fields, including optical object detection, mobile edge computing, voice assistants for smartphones, and Google keyboard query suggestion \cite{aledhari2020}. These applications, however, face many difficulties, including communication effectiveness, statistical and system heterogeneity, privacy, personalization, fairness etc. \cite{kairouz2021}. The elimination of demographic bias based on sensitive attributes of clients such as \textit{gender}, \textit{race}, \textit{age} etc., in Federated Recommendation Systems (FRSs) is the main theme of this paper. \par
\begin{figure}
\begin{subfigure}{1.0\linewidth}
  \centering
  \includegraphics[width=0.49\linewidth]{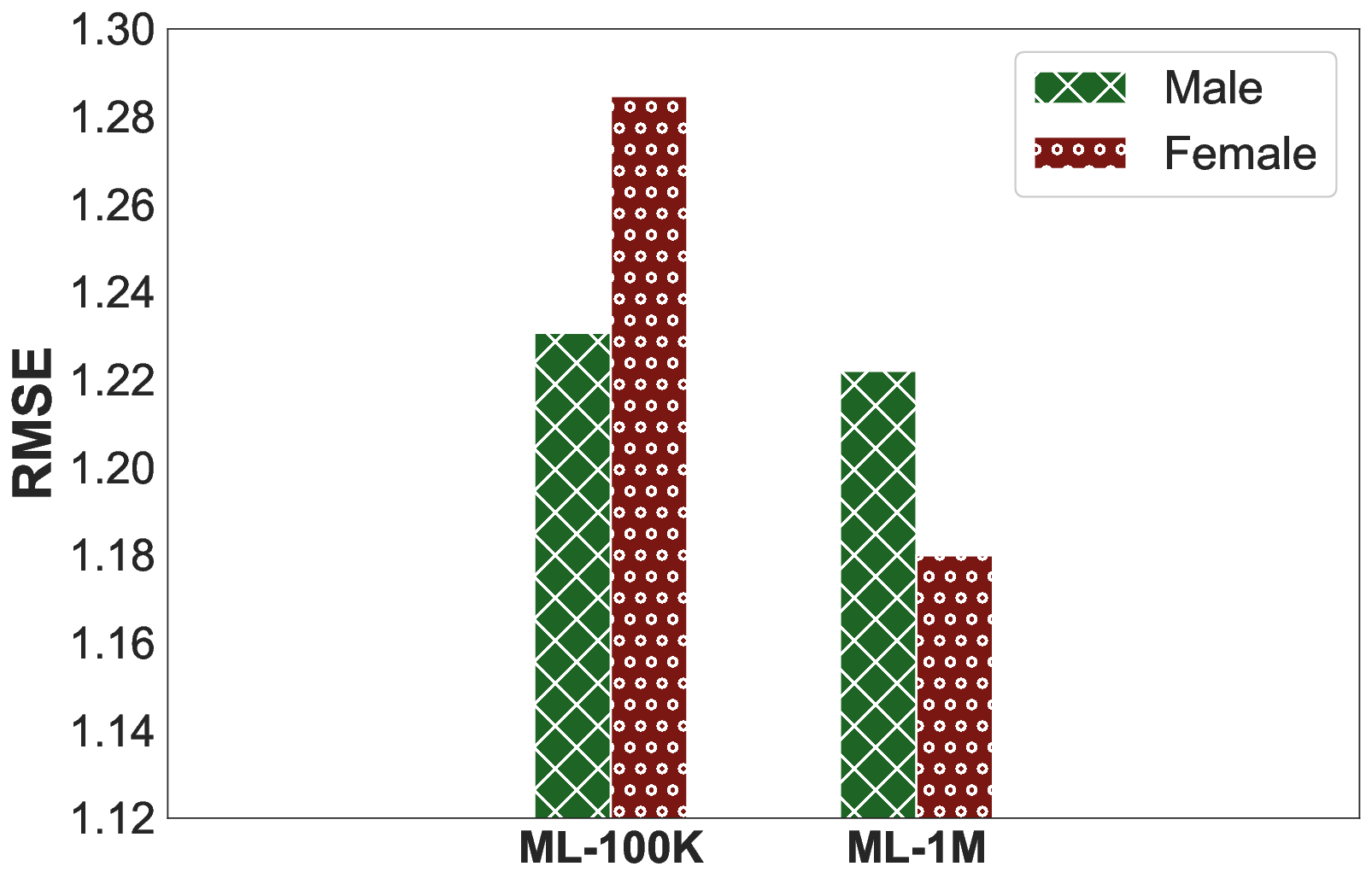}
  \includegraphics[width= 0.49\linewidth]{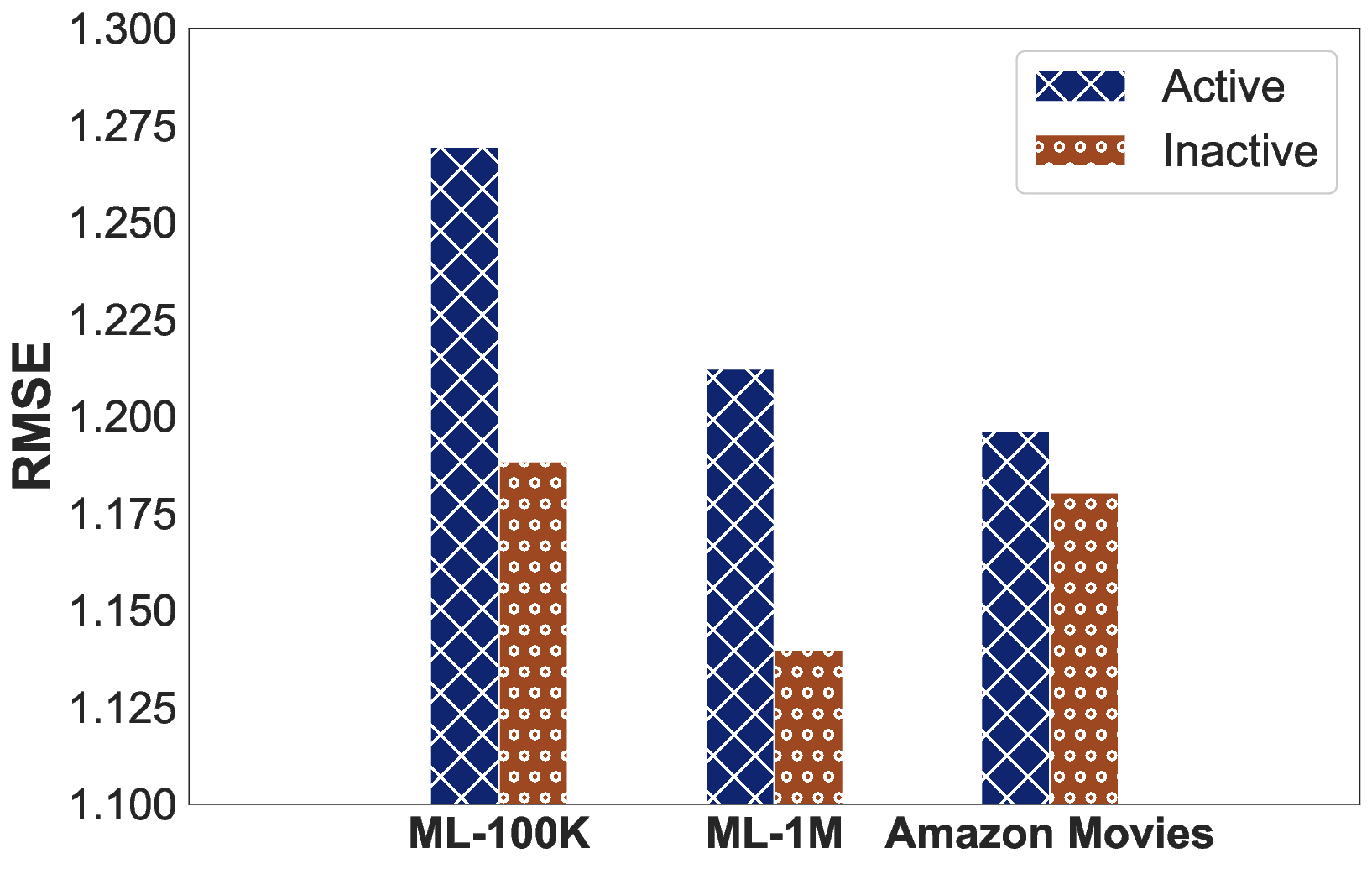}
\end{subfigure}
\caption{Disparity (group unfairness) in RMSE over two user attributes for F$^{2}$PGNN when fairness budget, $\beta = 0$: \textbf{(Left)} Gender \textbf{(Right)} Activity}
\label{fig:Originaldisparity}
\end{figure}
Many prior works \cite{islam2019mitigating, yao2017beyond, li2021user} mitigate unfairness in conventional RSs, which call for exchanging private attributes with the server and compromise privacy in federated settings. Contrary to this, training locally to achieve fairness in FRS without exposing user demographic information becomes exceptionally challenging. The backbone for existing FRS methods is Matrix Factorization (MF) \cite{chai2020secure}, in which the client updates user and item embeddings locally. In MF, only explicit user-item interactions are catered for updating embeddings; the implicit intricate interaction in the form of a bipartite graph does not play any role. To further utilize this graph structure in RSs data, the recent developments of Graph Neural Networks (GNNs) paved the way to build a GNN-based recommendation system \cite{wu2022graph}. Owing to its efficiency and inductive learning capability, GNNs for RSs are found superior compared to other approaches which are limited to transductive settings and cannot incorporate user attributes. GNNs can naturally encode implicit collaborative interactions along with explicit structure enforced to enhance user-item representation, resulting in improved recommendation quality. However, it has been found that GNN perpetuates biases in recommendations among the demographic groups  \cite{wu2022federated} (See Figure \ref{fig:Originaldisparity}). \par
Motivated by this finding, in this work, we propose F$^{2}$PGNN, a novel framework to train fair models for FRS (Figure \ref{fig:F2PGNNFramework}). In F$^{2}$PGNN, each client locally train a GNN after incorporating higher-order information into their local user-item subgraph utilizing our proposed Inductive Graph Expansion algorithm. Moreover, we encode fairness constraint to the objective function as a regularizer. This, in turn, requires a client to communicate only the demographic group statistics to the server in each FL round along with the model parameters after local updates. Next, the server aggregates this information to update the parameters and then re-broadcast it to each user, this process repeats until the convergence. To amplify the privacy protection in tandem with enhanced fairness, we additionally deploy Local Differential Privacy (LDP) \cite{Dwork14} to the model parameters and group statistics to ensure that the server remains unaware of any specific details regarding individual clients' datasets. F$^{2}$PGNN is the first framework, to the best of our knowledge, which focuses on fairness with the graph-based model in a federated setting. We present the details of F$^{2}$PGNN in Section \ref{sec:Main}.  Our significant contributions are summarized as follows:
\begin{itemize}
    \item We present a novel architecture, F$^{2}$PGNN, with three pillars of social benefit, namely Fairness, Privacy and Personalization for a recommendation system based on GNN in FL setting. 
    \item We introduce the Inductive Graph Expansion algorithm with privacy preservation, which minimizes communication overhead while effectively capturing higher-order interaction from distributed user data.
    \item To enhance privacy protection, we incorporate an additional LDP module for the model updates along with preserving the privacy of group statistics.
    \item Extensive experiments on three publicly available datasets (one small and two large-scale) elucidate the effectiveness of F$^{2}$PGNN. Detailed analysis and ablation study further validates the strength and efficacy of individual components proposed in F$^{2}$PGNN.
\end{itemize}
\section{Related Work} \label{sec:RelatedWork}
\textbf{Federated Graph Neural Networks in Recommendation:} GNNs have proven effective in modelling graph-structured data since they capture topological and higher-order information on graphs. In the context of recommendation, GNNs have shown promise in capturing complex relationships and dependencies among users, items and their interaction \cite{wang2019neural, he2020lightgcn, berg2017graph,ying2018graph}. Traditional GNN models for the recommendation need user data to be centralized to build a global graph representation. This, in turn, impedes the privacy of user data. However, data protection regulatory norms such as General Data Protection Regulation (GDPR) will restrict online platforms from storing user data centrally to learn a GNN model \cite{magdziarczyk2019right}. \par
To overcome this constraint, the Federated Graph Neural Networks (FedGraphNNs) concept has been proposed \cite{he2021fedgraphnn}. FedGraphNNs combines the strength of GNNs with FL, a privacy-preserving approach that permits collaborative model training across decentralized data sources without exposing raw data. Several works have explored FL for recommendation and privacy-preserving learning. FedMF \cite{chai2020secure} and Federated Collaborative Filtering (FCF) \cite{ammad2019federated} are two Matrix-Factorization based frameworks to learn user/item embeddings for RSs \cite{koren2009matrix}. As model updates can still reveal sensitive information, \cite{mcsherry2009differentially} proposed to use Differential Privacy to limit the exposure of user data. FedPerGNN \cite{wu2022federated} and FeSoG \cite{liu2022federated} are the most recent works that combine GNN with FRSs, which maximally aligns with our work. FedPerGNN uses repeated expensive encryption in each FL round; FeSoG takes only limited interactions as it considers only trusted users, which are the caveats of these algorithms. In addition, these works do not consider fairness constraints.\par
\noindent \textbf{Fair Federated Learning for Recommendation:}
Fair ML methods for RSs have been extensively explored in centralized settings compared to federated settings \cite{Wang2023, li2021tutorial, gao2022fair}. The availability of the whole dataset makes the application of existing fairness notions straightforward in centralized learning, whereas it is challenging to apply fairness in FL. Hence, different notions of fairness have been invented in FL, such as \textit{client-based fairness}, which enforces the parity across the clients, \textit{collaborative fairness}, which provide more reward to more contributing client \cite{wang2021federated}. There is another important fairness notion known as \textit{group fairness} in FL, in which each client belongs to a particular demographic group, and the group fair model does not discriminate against any group \cite{du2021fairness}.  \par
For FRSs, \cite{maeng2022towards} proposed a framework to model the interdependence of data and system heterogeneity. In \cite{Liu2022}, authors proposed a framework for a group fair FRS with privacy protection based on matrix factorization (F2MF). In F2MF, each client locally updates its embedding and does not consider the higher-order interaction resulting in more inherent unfairness. This is the only work in the literature for FRS with fairness and in line with our proposed F$^{2}$PGNN framework. Contrary to F2MF, F$^{2}$PGNN  take higher-order interaction into account when locally updating user and item embeddings. Also, to cater data heterogeneity across the clients,  F$^{2}$PGNN is personalized. In the next section, we give a brief background on fairness and GNN-based recommendation in the FL setting.             
\section{Background and Preliminaries} \label{sec:Background}
\textbf{FL with GNN based recommendation:}
GNN techniques have been demonstrated to be powerful for representation learning in RSs, as recommendation data inherently possess a graph-like structure. For instance, a bipartite graph connecting the user and item nodes can be used to represent the user-item interaction data, with each edge denoting a user-item interaction. In general, we can use any local GNN architecture, viz. GCN \cite{kipf2017}, GraphSAGE \cite{graphsage}, GAT \cite{gat2018} etc. In this paper, we have adopted GAT architecture for local GNN to learn user and item embeddings while investigation with others is straightforward. \par
To formulate the FL setting, we define $\Theta$ as the overall learnable weights for the GNN of user $u$. Hence, FL with GNN can be formulated as distributed optimization problem following the standard FedAvg \cite{mcmahan2017} algorithm as follows:
\begin{equation}
    \min_{\Theta}f(\Theta) = \min_{\Theta} \sum_{u=1}^{N} p_{u} \cdot \mathcal{L}_u (\boldsymbol{\Theta})
\end{equation}
\noindent where $\mathcal{L}_u(\boldsymbol{\Theta})=\frac{1}{N_u} \sum_{(x,y) \in \mathcal{G}_u} \mathcal{L}\left(\boldsymbol{\Theta}, x, y\right)$ is the local objective of user $u$ that measures empirical risk over local dataset $\mathcal{G}_u$ of size $N_u$; $p_u \geq 0$ and $\sum_{u} p_u=  1$. The loss function for the global GNN model is $\mathcal{L}$. Here, GAT employs an attention mechanism to distinguish the importance of neighbouring nodes and updates the embedding of each node by attending to its neighbours as in Eq.(\ref{eq:GNN_equations}).
\begin{equation} \label{eq:GNN_equations}
\resizebox{.3 \textwidth}{!}
{$
\begin{aligned}
      & Aggregation: \quad \mathbf{n}_v^{(l)}=\sum_{k \in \mathcal{N}_v} \gamma_{v k} \mathbf{h}_k^{(l)},\\
      & \quad  \gamma_{v k}=\frac{\exp \left(\operatorname{Att}\left(\mathbf{h}_v^{(l)}, \mathbf{h}_k^{(l)}\right)\right)}{\sum_{j \in \mathcal{N}_v} \exp \left(\operatorname{Att}\left(\mathbf{h}_v^{(l)}, \mathbf{h}_j^{(l)}\right)\right)}, \\
      & \quad Update: \quad \mathbf{h}_v^{(l+1)}=\sigma\left(\mathbf{\Theta}^{(l)} \mathbf{n}_v^{(l)}\right)
\end{aligned}
$}
\end{equation}
\noindent where $\mathbf{h}_v^{(l)}$ indicates representation of node $v$ at $l^{th}$ layer and $\mathcal{N}_v$ is the set of neighborhood of node $v$. $\operatorname{Att}\left(\cdot\right)$ is the attention function and typically $\operatorname{Att}\left(\cdot\right)$ is $\operatorname{LeakyReLU}\left(\mathbf{a}^T\left[\mathbf{\Theta}^{(l)} \mathbf{h}_v^{(l)} \oplus \mathbf{\Theta}^{(l)} \mathbf{h}_k^{(l)}\right]\right)$, $\mathbf{a}$ is the learnable parameter and $\mathbf{\Theta}^{(l)}$ are the model parameter for transforming node representation at $l^{th}$ layer, $\oplus$ is the concatenation operation, $\sigma\left(\cdot\right)$ is the non-linear activation function. 
 In the following subsection, we outline the fundamental notions of fairness in ML.\par
\vspace{0.3cm}
\noindent \textbf{Notion of Fairness:}
In critical ML applications, if data of individuals contain sensitive demographic information such as gender, race etc., then the trained model may have discriminatory outcomes based on this sensitive attribute. For such models, the fairness is evaluated with respect to its performance compared to the underlying groups defined by the sensitive attribute  $S$, and this notion of fairness is called \textit{group fairness}. For a sensitive attribute $S$, if the privileged group (e.g., male) is denoted by $S = 1$ and $S=0$ denotes the underprivileged group (e.g., female), the model's prediction for the positive class is assumed to be positive. In such a scenario, how the prediction of the model is being considered, several notions of group fairness have been proposed in the literature for centralized training, viz. \textit{Equalized Odds}, \textit{Equality of Opportunity} \cite{Hardt2016} and \textit{Statistical Parity} \cite{Dwork2012} etc. 

The above-mentioned notions of fairness are applicable in centralized ML algorithms because data and information of sensitive attribute is available. Based on the $S$ dataset is divided into subgroups, and the desired metric can be calculated. However, in FL settings, without accessing the client's sensitive attributes, the server cannot apply these fair centralized ML techniques, which require this information on a global level to achieve fair classification. Hence, to extend group fairness to FL, we should devise a fairness metric applicable in the FL setting. We define user-level fairness as follows:
\begin{definition}
Let the recommendation list for user $u$ is denoted by $\mathcal{R}_{u}$, for a given performance evaluation metric $\mathcal{M}$, the group fairness for $u$ with respect to groups, $S_{0}$ and $S_{1}$ is defined as
$$ \mathbb{E}_{u}[\  \mathcal{M}(\mathcal{R}_{u})| u \in S_{0} ]\ = \mathbb{E}_{u}[\  \mathcal{M}(\mathcal{R}_{u})| u \in S_{1} ]\ . $$
\end{definition}
Also, the \textit{Equalized Odds} notion of fairness amounts to the mistreatment of groups and can be interpreted as gap between group-average performance. Hence, the group (un)fairness for two mutually exclusive groups of users in the FL setting can empirically be measured as
\begin{equation} \label{Eq:disp_def}
\resizebox{.43 \textwidth}{!}
{$ \mathcal{L}_{fair}(\mathcal{M}, S_{0}, S_{1}) = \left|\frac{1}{|S_{0}|} \sum_{u \in S_{0}} \mathcal{M}(u) - \frac{1}{|S_{1}|} \sum_{u \in S_{1}} \mathcal{M}(u)\right|^{\alpha} $} 
\end{equation}
where, $\alpha$ determines the smoothness and can take integer values $1$ or $2$ (we set $\alpha = 1$). Here, the evaluation metric $\mathcal{M}$, determines the performance of each user, and Eq.(\ref{Eq:disp_def}) quantifies the global (un)fairness of the model. The small value of $\mathcal{L}_{fair}$ indicates the model is fair, and minimization of this term while maintaining the model efficacy becomes the ultimate goal to achieve fair recommendation model. In Eq.(\ref{Eq:disp_def}), we have given $\mathcal{L}_{fair}$ formulation for binary sensitive attribute, the extension of this to multi-group is straightforward \cite{Liu2022}. 
\begin{figure*}[h]
    \centering
    \includegraphics[width = \textwidth, height=5.5cm]{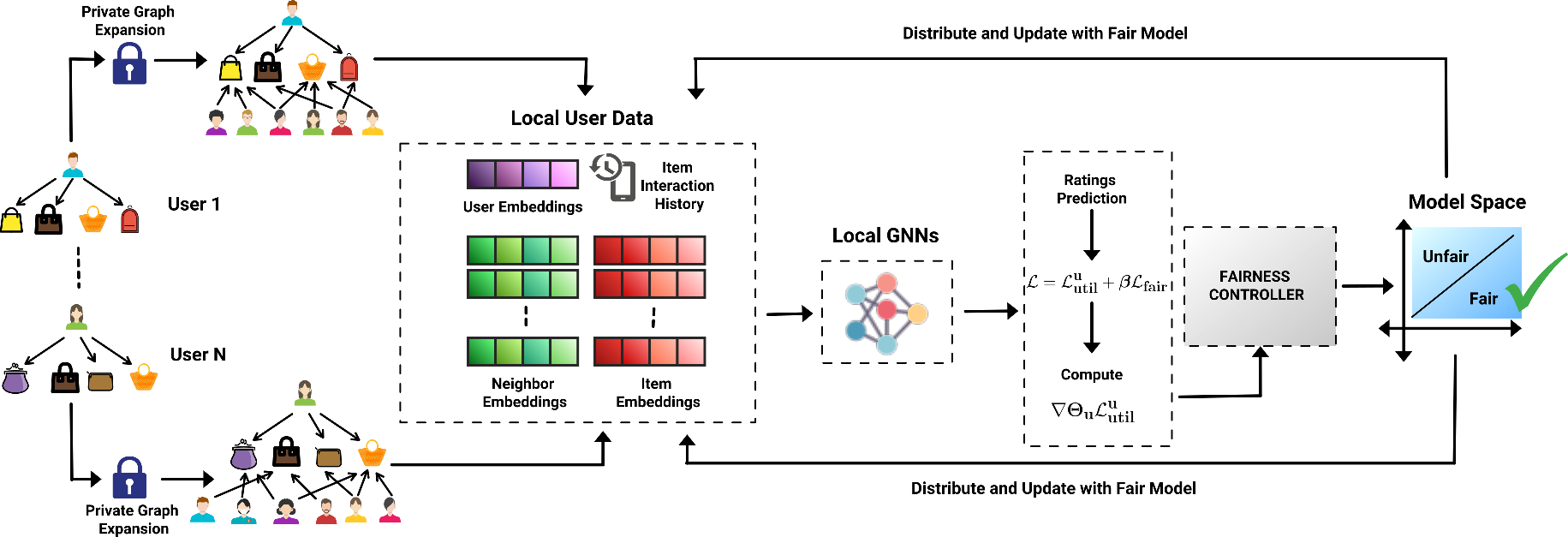}
    \caption{{\scshape F$^{2}$pgnn}: Schematic of Group-Fairness aware Federated Graph Neural Network for Personalized Recommendation.}
    \label{fig:F2PGNNFramework}
\end{figure*}
\section{F$^{2}$PGNN: Fairness aware GNN in FL} \label{sec:Main}
In this section, we present the details of F$^{2}$PGNN, an approach to introduce fairness in FRSs with graphical models, and finally, we analyze privacy protection while achieving global group fairness.
\subsection{F$^{2}$PGNN framework} \label{sec:ProposedFramework}
As in general federated settings, each user in F$^{2}$PGNN framework stores its user-item interaction history to constitute a local subgraph. To train a personalized GNN, first, each user expands the local subgraph using the Inductive Private Graph Expansion algorithm to incorporate higher-order interactions (Appendix \ref{IG_expension}-Algorithm \ref{alg:privategraphalgorithm}). By matching the encrypted items and distributing anonymous user embeddings, the extended graph includes the neighbors of each user with co-interacted items. For each user $u_{i}$ that has interacted with $m$ items and $r$ neighbors with co-interacted items, an embedding layer is used to create it's node embedding $z_{i}^{u}$, item embeddings $\left[ z_{i,1}^{t}, z_{i,2}^{t}, \cdots z_{i,m}^{t} \right]$ and embeddings of neighbors $\left[ z_{i,1}^{u}, z_{i,2}^{u}, \cdots z_{i,r}^{u} \right]$. Next, a GAT model is used to update these embeddings based on the local subgraph. The final representation of GAT model for user and item nodes are denoted as $h_{i}^{u}$, $\left[ h_{i,1}^{t}, h_{i,2}^{t}, \cdots h_{i,m}^{t} \right]$ and $\left[ h_{i,1}^{u}, h_{i,2}^{u}, \cdots h_{i,r}^{u} \right]$ for the prediction task.
Then, for each user $u_{i}$, the recommendation loss will be $\mathcal{L}_{util}^{u} = \frac{1}{m}\sum_{j=1}^{m}|\hat{y}_{i,j}- y_{i,j}|^2$ where, $\hat{y}_{i,j}$,  $y_{i,j}$ denotes the predicted rating and true rating respectively, for user $i$ and item $j$. The loss $\mathcal{L}_{util}^{u}$ is used to obtain the local gradient $ \nabla \Theta _{u}$ of the model and the performance for each user. \par
Now, to incorporate global fairness in this recommendation model, we formulated a combined optimization problem. For any $\beta \in \left[0,1\right)$, we have:
\begin{equation}\label{eq:losstotal}
\mathcal{L} = \mathcal{L}_{util} + \beta \mathcal{L}_{fair}
\end{equation} 
where $\beta$ is a hyperparameter that denotes the trade-off between utility and fairness, and $\mathcal{L}_{util} = \frac{1}{N} \sum_{u=1}^{N} \mathcal{L}_{util}^{u} $, $\mathcal{L}_{fair}$ is given in Eq.(\ref{Eq:disp_def}). It is important to note that, for calculating $\mathcal{L}_{fair}$ at the server, the group information of each user is required which is in contradiction with the principles of FL. To overcome this conflicting situation, we have extended the privacy-preserving mechanism used in \cite{Liu2022}; an overview of the framework is shown in Figure \ref{fig:F2PGNNFramework}.     

\subsubsection{Strategy for optimizing the global loss function privately:} The problem in Eq.(\ref{eq:losstotal}) can be effectively minimized using the stochastic gradient descent method if it is differentiable. We consider $\mathcal{M}_{u} = - \mathcal{L}_{util}^{u}$ as a measure of performance for each user $u$, then local loss function becomes differentiable and suitable for FL process as shown below. Let, the gradient of the Eq.(\ref{eq:losstotal}) for any user is,       
\begin{equation}\label{gradeq}
    \nabla \Theta _{u} = \frac{\partial}{\partial \Theta_{u}}\mathcal{L}_{util}^{u} + \beta \frac{\partial}{\partial \Theta_{u}}  \mathcal{L}_{fair}
\end{equation}
To simplify the above expression, let $P\!= \! \frac{1}{|S_{0}|} \! \sum_{u \in S_{0}} \!\mathcal{M}_{u}$  and $Q\! = \frac{1}{|S_{1}|} \sum_{u \in S_{1}} \mathcal{M}_{u}$ be the global group statistics ($\mathcal{G}_{stat}$). Since, $\mathcal{M}_{u} = - \mathcal{L}_{util}^{u}$, then from Eq.(\ref{Eq:disp_def}), for each user we have,
\begin{equation}\label{gradfair}
    \frac{\partial}{\partial \Theta_{u}}  \mathcal{L}_{fair} = -R\left | P - Q \right |^{\alpha - 1} \frac{\partial}{\partial \Theta_{u}}\mathcal{L}_{util}^{u}
\end{equation}
Now, combining the Eq.(\ref{gradeq}) and Eq.(\ref{gradfair}) we have,
\begin{equation} \label{eq:scale}
\begin{split}
 \nabla \Theta _{u} \!\!& = \! \! \left(\!1 \!-\! \beta R\left | P\! -\! Q \right |^{\alpha - 1}\! \right) \! \frac{\partial}{\partial \Theta_{u}}\mathcal{L}_{util}^{u}
 \!= \!L \frac{\partial}{\partial \Theta_{u}}\mathcal{L}_{util}^{u}
\end{split}
\end{equation}
where, $L \! = \!1 \!-\! \beta R\left | P - Q \right |^{\alpha - 1}$. Here, $R = \alpha(-1)^{\mathds{1}(P < Q)} (-1)^{\mathds{1}(u \notin S_{0})}$, hence for $R \!\! >\!\! 0$, user $u$ belongs to the superior performance group, thus $L < 1$, which slows down the learning of the user $u$,  otherwise $R \!\leq 0 \!\Rightarrow \! L \geq 1$ which scales up the learning for poor performing user. Hence, fairness can be achieved by regulating the learning rates. Similarly, this mechanism can be extended to multi-group scenarios in a federated setting.
Thus, enforcing fairness for each user becomes straightforward with a small overhead in communication. Further, the key benefit of this fairness algorithm is its model-agnostic nature which perfectly fits in federated settings. The schematic diagram is presented in Appendix \ref{FC} (Figure \ref{fig:Fairnesscontroller}). 

\subsection{Privacy Protection in F$^{2}$PGNN} \label{sec:PrivacyProtection}
The user's privacy in F$^{2}$PGNN is safeguarded through three key aspects. \par
   \noindent $\bullet$ \textbf{Secure user-item local graph expansion:} In GNN-based RSs, getting higher-order interactions without violating user privacy in FL settings becomes challenging. FedPerGNN \cite{wu2022federated} handles this using repeated expensive encryption. To overcome this bottleneck, we developed an inductive user-item graph expansion algorithm in a privacy-preserving manner; the pseudocode for algorithm is given in Appendix \ref{IG_expension} (Algorithm \ref{alg:privategraphalgorithm}). Each user encrypts the items using the public key and uploads the encrypted IDs to the server. After matching the encrypted items, the server then distributes anonymous user embeddings to each user for expanding their local subgraph. Moreover, as the server has access to the previously rated encrypted item IDs for each user, only the newly rated items need to be encrypted in future communication rounds. That makes this algorithm inductive in nature, in addition to reducing the communication overhead. \par 
    \noindent $\bullet$ \textbf{Privacy preserving model update:} We used the standard LDP technique to overcome the privacy leakage of user-item interaction history if a user directly uploads the model parameters \cite{choi2018guaranteeing} (Refer Appendix \ref{ldp} for more details on LDP).  Following the standard procedure, we clip the scaled gradients based on their $L_{2}$-norm with a clipping threshold $\delta$, and add zero-mean Laplace noise with $\lambda$ variance to obtain $\epsilon$-LDP. To control the amount of noise and clip, the upper bound for the privacy budget $\epsilon$ is $\frac{2\delta}{\lambda}$ \cite{qi-etal-2020}. After protecting gradients, the user performs a local update and uploads the updated model parameters to the server for aggregation; in the interest of space, we defer the pseudocode for the algorithm in Appendix \ref{IG_expension} (Algorithm \ref{alg:localupdate}). \par       
     \noindent $\bullet$ \textbf{Secure group statistics aggregation:} For updating the global model, the server requires the group statistics, $P$ and $Q$, i.e the information about whether user belongs to $S_{0}$ or $S_{1}$. Uploading information about the user's sensitive attributes violates the user's privacy in FL. We aggregate group statistics in a secure manner using LDP \cite{Liu2022}, in which first each user $u$ uploads $\mathcal{G}_{stat}^{u}$ as follows: 
    \begin{equation} \label{eq:g_stats} 
    \resizebox{0.40 \textwidth}{!}
    {$
    \begin{split}
    P_{per}^{u}  \! & \leftarrow  \!\mathds{1}(u\! \in \!S_{0})\mathcal{M}_{u} \! \!+\! \epsilon_{1,u},
    P_{add}^{u} \! \leftarrow \!\mathds{1}(u\! \in\! S_{0}) \!+\! \epsilon_{3,u} \\ 
    Q_{per}^{u} \! & \leftarrow \!\mathds{1}(u\! \in \! S_{1})\mathcal{M}_{u} \! \! + \! \epsilon_{2,u},  
    Q_{add}^{u} \! \leftarrow \! \mathds{1}(u\! \in \! S_{1}) \!+\! \epsilon_{4,u}\\
    \end{split}
    $}
    \end{equation}
    where, $\epsilon_{1,u},\epsilon_{2,u}, \epsilon_{3,u}, \epsilon_{4,u} \sim \mathcal{N}(0, \sigma^{2})$ are personalised noise fixed for each epoch.  Then the server can aggregate $\mathcal{G}_{stat}^{u}$ as follows,   
    \begin{equation} \label{eq:agg}
        P = \frac{\sum_{u} P_{per}^{u}} {\sum_{u} P_{add}^{u}} , \text{and }  Q = \frac{\sum_{u} Q_{per}^{u}}{ \sum_{u} Q_{add}^{u}}.
    \end{equation}
The detailed pseudo-code of F$^{2}$PGNN algorithm is given in Algorithm \ref{alg:algorithm}.
\begin{algorithm}[!h]
\caption{\textbf{F$^{2}$PGNN Algorithm}}
\label{alg:algorithm}
\textbf{Input}: Initialize local subgraphs $\mathcal{G}_{u}$, global model $\Theta^{0}$ \\
\textbf{Parameter}: Learning rate $\left(\eta\right)$, Noise parameter $\left(\sigma\right)$, Batch Dropout rate $\left(K\right)$, Hidden dimension $\left(h\right)$,  Fairness Budget $\left(\beta \right)$, Group Statistics $\left(\mathcal{G}_{stat}\right)$ : $P^{0}, Q^{0} \leftarrow 1$\\
\textbf{Output}: Fair Model $\Theta$, User Embeddings, Item Embeddings
\begin{algorithmic}[1] 
\STATE Randomly sample set of users $U_{K}$ with $ \left | U_K \right | = \left ( 1-K \right )\cdot \left | U \right |$, where $\left|U\right| = $ Total number of users  
\WHILE{not converged in epoch $i$}
\STATE Broadcast $\Theta^{i}$ \& $\mathcal{G}_{stat}$ to each user
\STATE // \textbf{User}
\FOR{each user $u \in U_K$}
\STATE Mapping-Dict $\leftarrow$ \textbf{PrivateGraphExpansion$\left(\right)$}
\STATE Expand $\mathcal{G}_{u}$ using Mapping-Dict
\STATE $\mathcal{L}_{util}^{u}$, $\nabla_{\Theta}\mathcal{L}_{util}^{u}$ $\leftarrow$ Local GNN training
\STATE Scale $\nabla_{\Theta}\mathcal{L}_{util}^{u}$ as per \eqref{eq:scale} \& update $\mathcal{G}_{stat}^{\left(u\right)}$ as per \eqref{eq:g_stats}
\STATE $\Theta_{u}^{i}$ $\leftarrow$ \textbf{LocalUpdate$\left(\right)$}
\STATE Upload $\Theta_{u}^{i}$ \& $\mathcal{G}_{stat}^{\left(u\right)}$ to the server
\ENDFOR
\STATE // \textbf{Server}
\STATE $\Theta^{\left ( i+1 \right )} \leftarrow AGG\left ( \Theta_{u}^{i} \bigg\rvert \forall u \in U_K \right)$
\STATE Update $\mathcal{G}_{stat}$ as per \eqref{eq:agg}
\ENDWHILE
\end{algorithmic}
\end{algorithm}

\section{Experimental Evaluation} \label{sec:Evaluation}
This section assesses the efficacy of F$^{2}$PGNN across different system settings. In particular, we evaluate how the trade-off between fairness, privacy and utility is influenced by the fairness budget $\beta$ and LDP parameters. 
\subsection{Experimental Setup} \label{sec:expsetup}
\subsubsection{Implementation:} We implemented F$^{2}$PGNN in Python 3.9 using TensorFlow 2.5. All experiments are performed on a machine with AMD EPYC 7282 16-Core Processor @ 2.80GHz with 128GB RAM, 80GB A100 GPU on Linux Server. The source code is given in supplementary material. \par 
\setlength{\fboxsep}{2pt}
\begin{table*}[h]
\centering
\begin{tabular}{p{5.5em} p{4.2em} p{6.2em} p{6.2em} p{6.2em} p{6.2em} p{6.2em}}
 \hline 
 \hline
 Dataset & Method  &  \multicolumn{5}{c}{ \centering RMSE(Disparity)} \\
   &\centering $\downarrow$  $\mathbf{\beta} \rightarrow$ & $\mathbf{0.0}$ & $\mathbf{0.3}$ & $\mathbf{0.5}$ & $\mathbf{0.7}$ & $\mathbf{0.9}$ \\ 
   \hline
\multirow{2}*{ML-100K(G)} & F2MF & 1.5801(0.1242) & 1.4905(0.1145) & \fbox{1.4569(0.0938)} & 1.4183(0.1169) & 1.5118(0.1213) \\
&F$^{2}$PGNN & \bf 1.2444(0.0539) & \bf 1.2545(0.0522)&\bf 1.2616(0.0512)&\bf 1.2686(0.0501)& \bf 1.2758(0.0491)\\
\multirow{2}*{ML-100K(A)} & F2MF & 1.5397(0.2452)& \fbox{1.4822(0.1777)} & 1.4625(0.2526) & 1.4305(0.1577) & 1.4966(0.2290) \\
& F$^{2}$PGNN& \bf 1.2478(0.0811) & \bf 1.2471(0.0806) & \bf 1.2467(0.0804) & \fbox{\bf 1.2466(0.0801)}& \bf1.2470(0.0804) \\
\hline
\multirow{2}*{ML-1M(G)} & F2MF & 1.2425(0.2441) & 1.2199(0.2391) & \fbox{1.1995(0.2327)} & 1.2026(0.2520) & 1.2236(0.2396) \\
&F$^{2}$PGNN & \bf 1.2118(0.0420) & \bf 1.2019(0.0410) & \bf 1.1950(0.0392) & \bf 1.1885(0.0380) & \bf 1.1831(0.0367)\\
\multirow{2}*{ML-1M(A)} & F2MF & 1.2663(0.8797) & 1.2161(0.8563) & 1.2101(0.8351)& 1.1872(0.8295)& \fbox{1.1773(0.7910)}\\
& F$^{2}$PGNN& \bf 1.1927(0.0724)& \bf 1.1843(0.0638)& \bf 1.1779(0.0567)& \bf 1.1711(0.0482) & \bf 1.1641(0.0379) \\
\hline
\multirow{2}*{\parbox{1.2cm}{\centering Amazon-Movies(A)}} & F2MF & \textbf{1.0968}(2.5444) & \textbf{1.0863}(2.5132) &\fbox{\textbf{1.0660}(2.4661)}& \textbf{1.0864}(2.5102) & \textbf{1.0907}(2.5157) \\
& F$^{2}$PGNN & 1.1889\bf(0.0157) & 1.1863\bf(0.0133) & 1.1843\bf(0.0114) & 1.1819\bf(0.0088) & 1.1788\bf(0.0052)\\
\hline
\hline
\end{tabular}
\caption{Performance vs Fairness comparison with different fairness budget $\beta$. For the user attributes in the dataset, G denotes Gender while A denotes Activity. Superior performing values are highlighted in bold. Box indicates the threshold on $\beta$.}
\label{Table:TestResults}
\end{table*} 
\subsubsection{Dataset:} To empirically evaluate our framework, we have used three publicly available real-world datasets, namely MovieLens (ML-100K and ML-1M versions) \cite{Harper15}, and Amazon-Movies ($\sim$500K ratings) \cite{ni-etal-2019}. We defer the summary statistics of these datasets to Appendix \ref{dataset_summary}. For all datasets, we first filter 20-core data \footnote[1]{https://github.com/CharlieMat/FedFairRec.git}, which ensures that each user has rated at least 20 items and each item has been interacted by at least 20 users, more details on the n-core dataset is given in Appendix \ref{n-core}. We then follow $80/10/10$ train/validation/test split for each user history sorted according to rating timestamps. We first consider the gender (G) of users as a sensitive attribute for ML-100K and ML-1M datasets, and we also include a synthetic attribute i.e activity (A) of users for all three datasets similar to \cite{Yunqi2021}, which considers users as \textit{active} if the number of ratings given by users exceed a certain threshold. \par

\subsubsection{Baselines:} The following is the only state-of-the-art FRSs method with fairness which is considered as baseline:
\begin{itemize}
    \item \textbf{F2MF} \cite{Liu2022} : A Matrix Factorization based approach in federated setting. It achieves fairness over different user demographic groups without exposing the sensitive user attribute.
\end{itemize} \par
\begin{figure*}[h]
\centering 
\includegraphics[width=.2\textwidth]{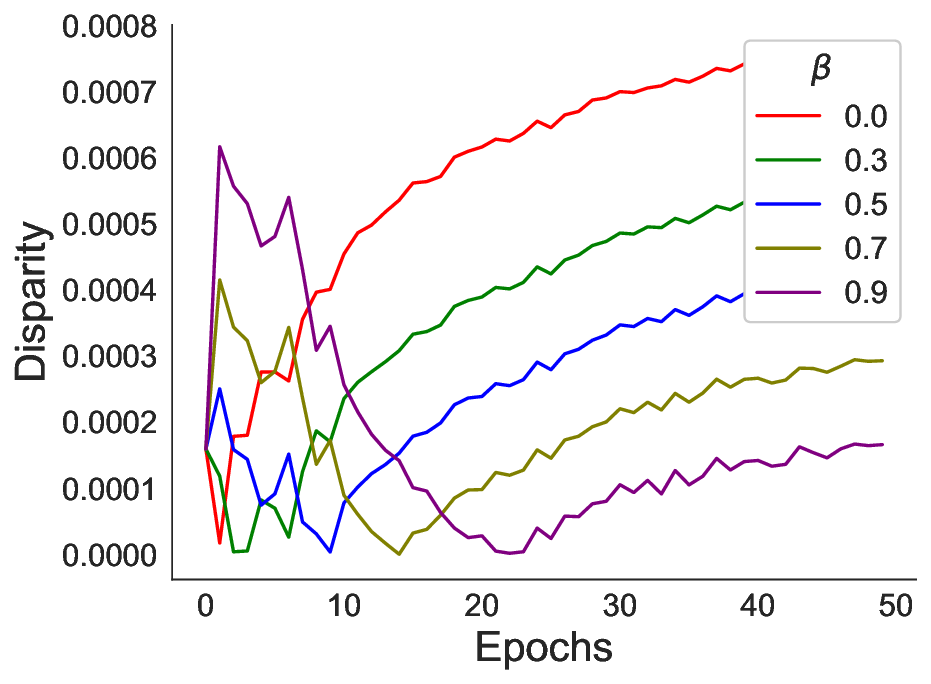}
\includegraphics[width=.2\textwidth]{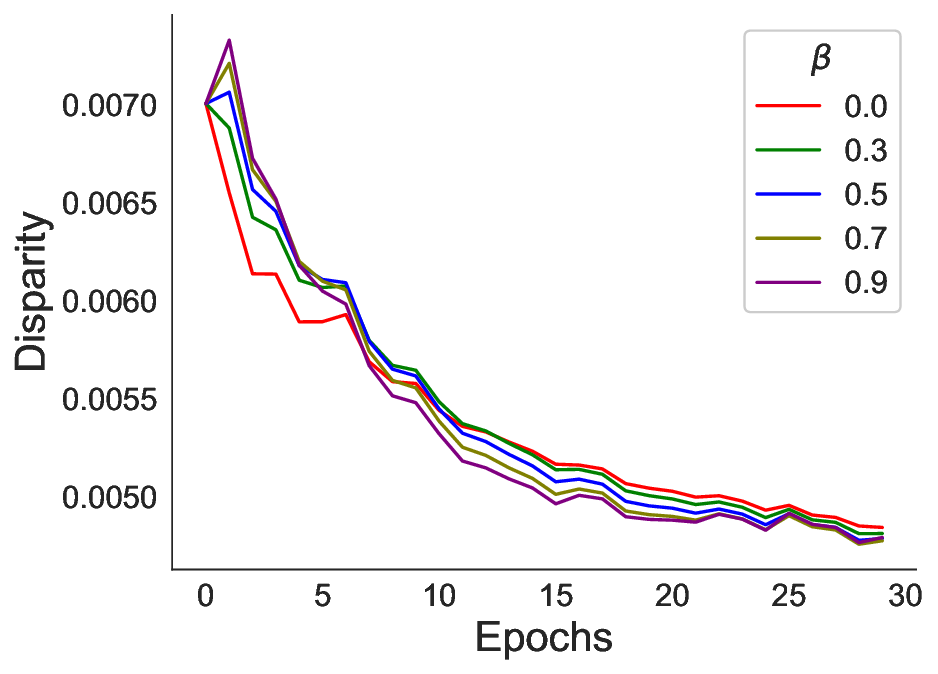}
\includegraphics[width=.2\textwidth]{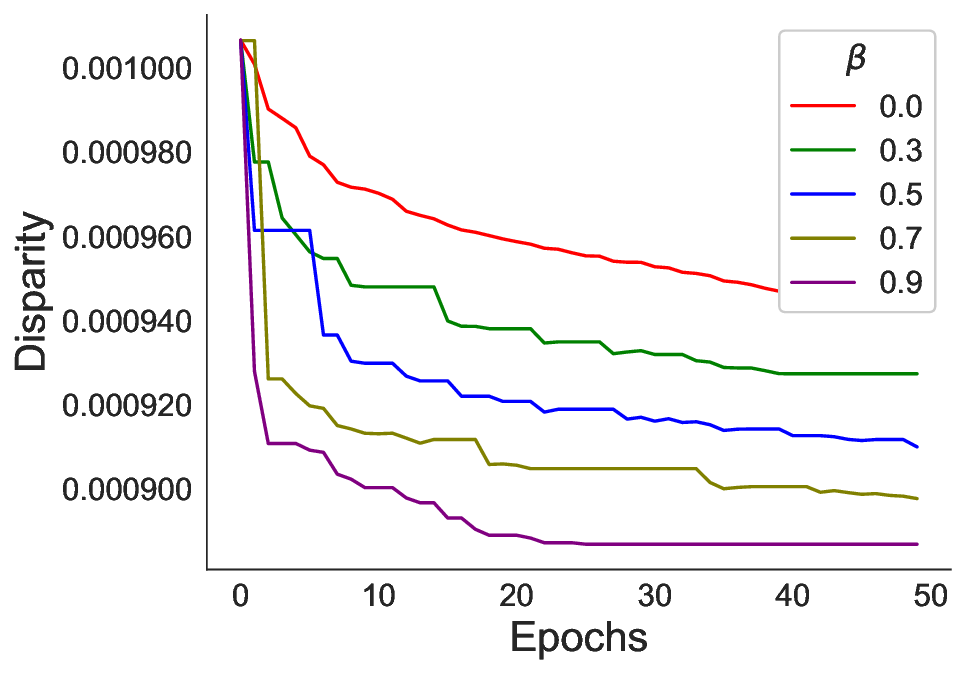}
\includegraphics[width=.2\textwidth]{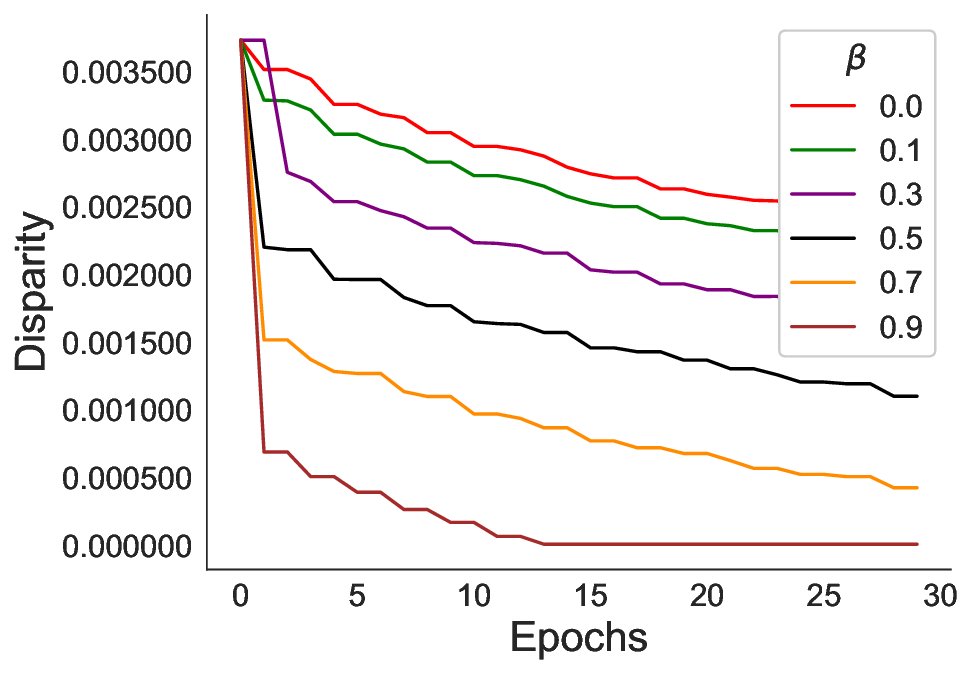}
\includegraphics[width=.2\textwidth]{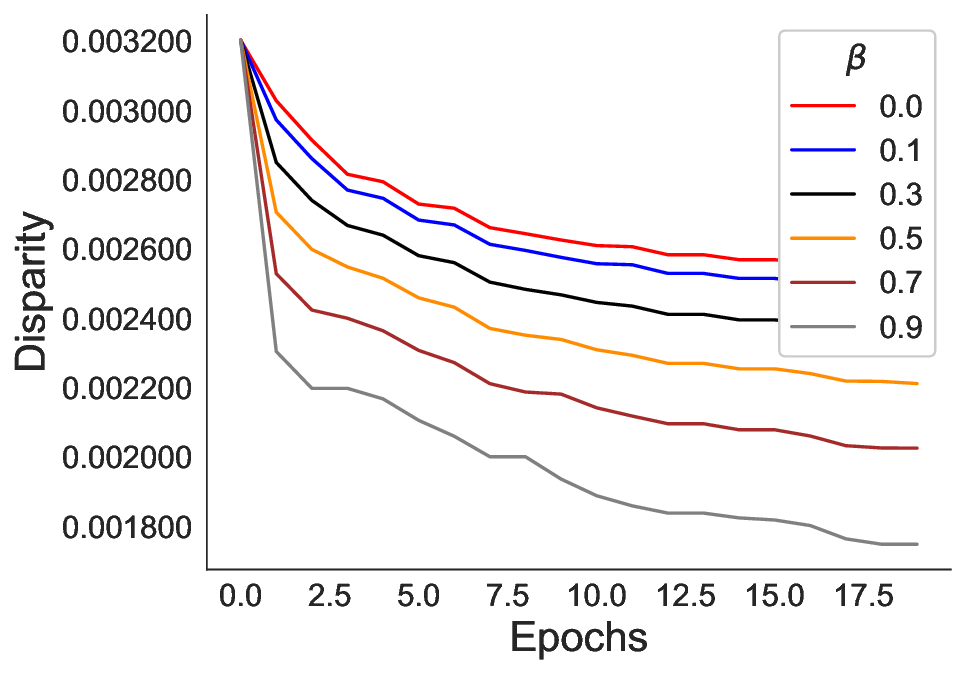}
\medskip
\includegraphics[width=.2\textwidth]{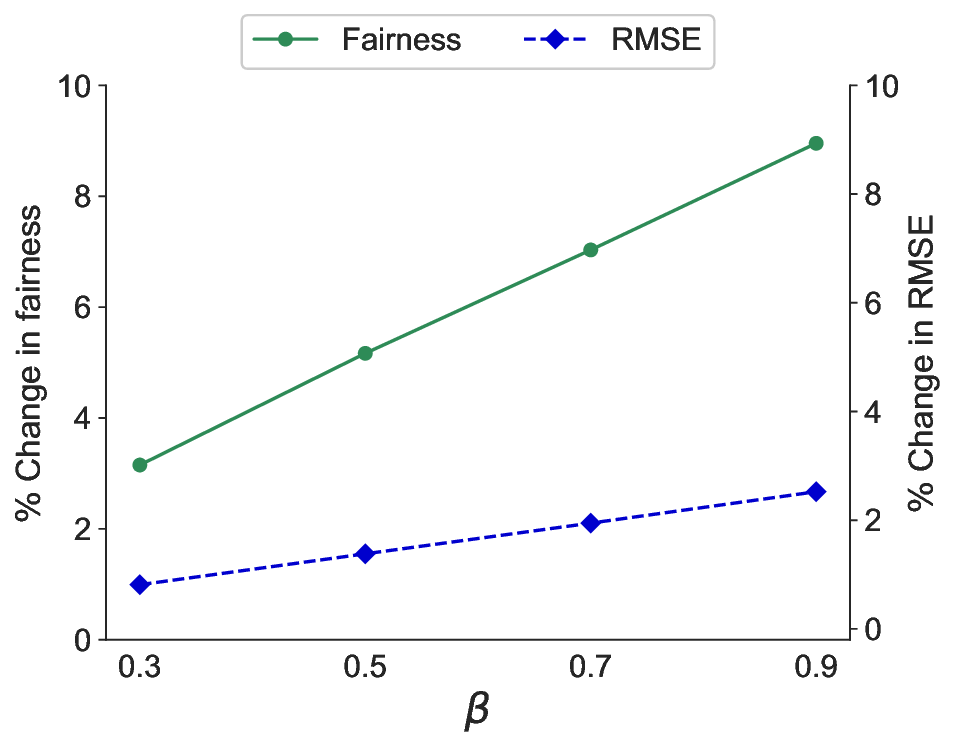}
\includegraphics[width=.2\textwidth]{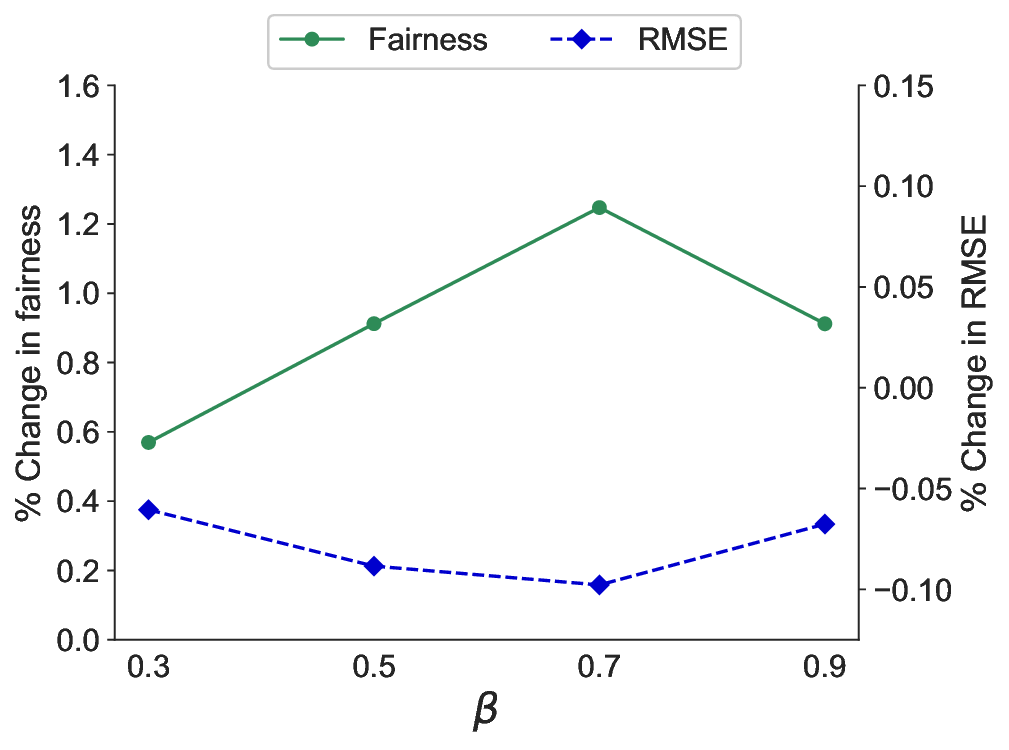}
\includegraphics[width=.2\textwidth]{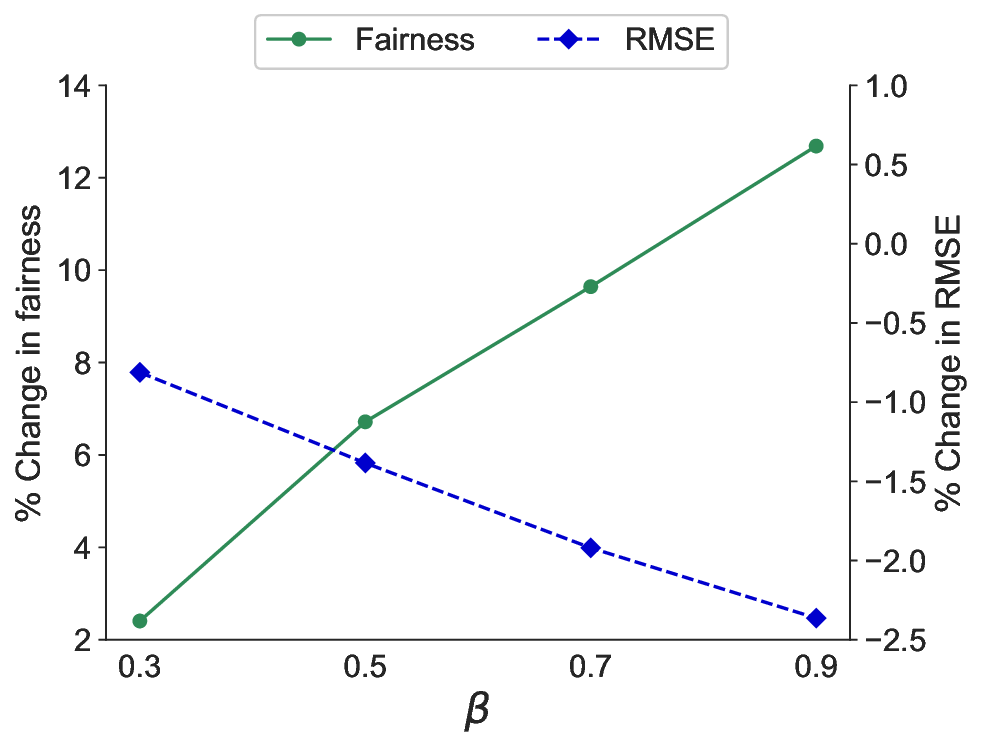}
\includegraphics[width=.2\textwidth]{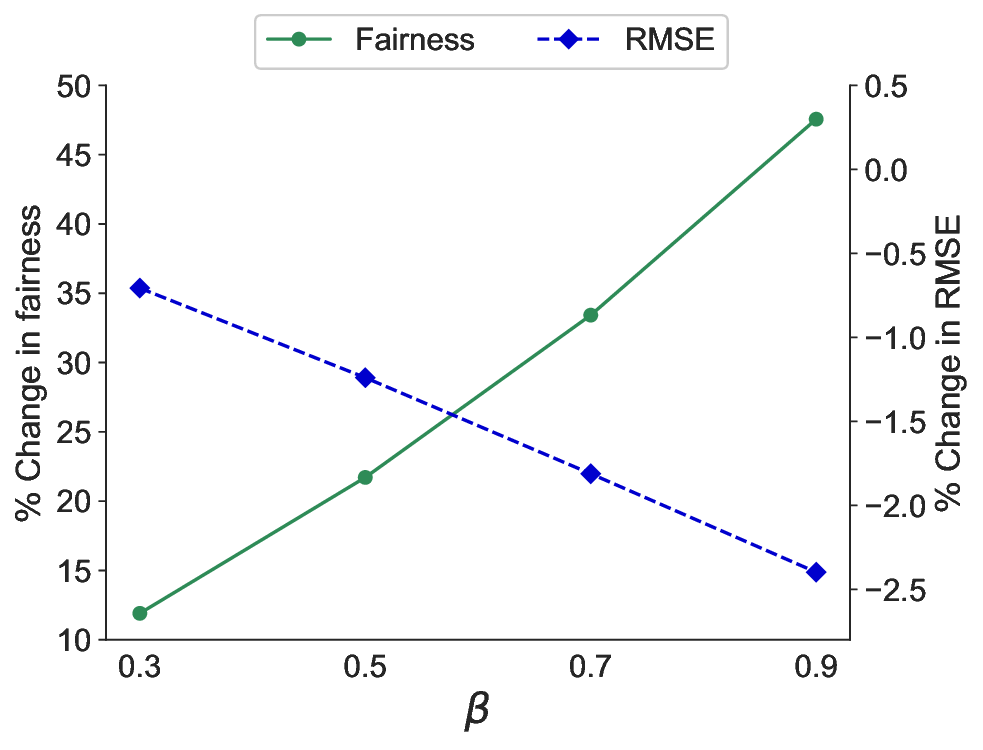}
\includegraphics[width=.2\textwidth]{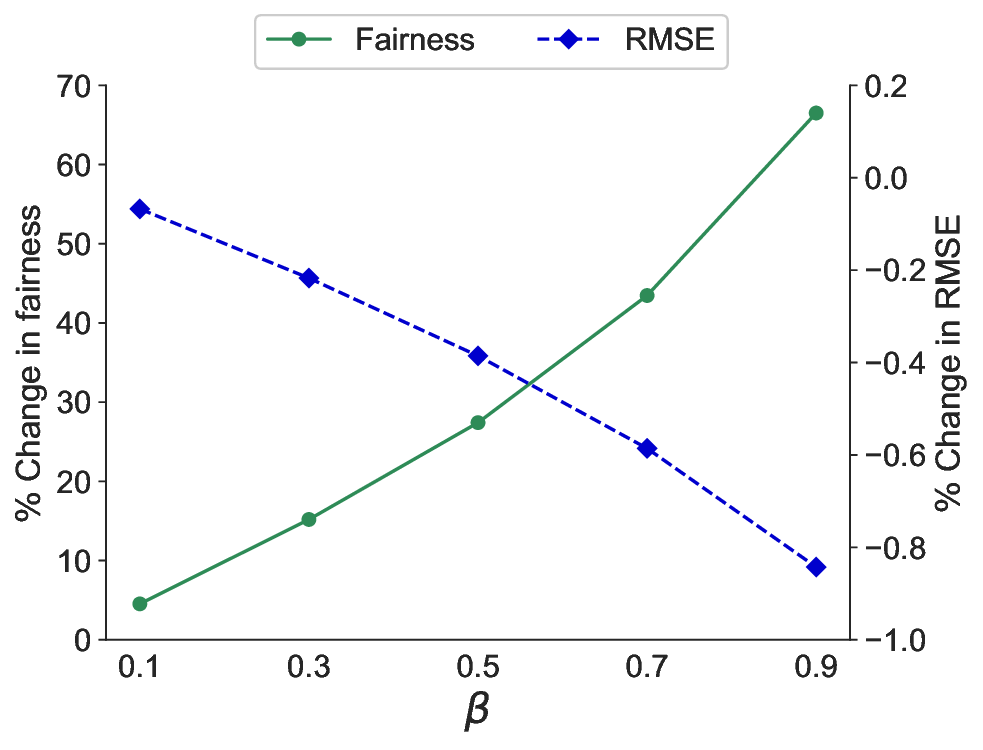}
\caption{\textbf{Top Row:} Disparity vs epoch for different fairness budget $\beta$ on validation data. Curves become lower with increasing $\beta$. \textbf{Bottom row:} $\%$ change in fairness (left y-axis) and $\%$ change in RMSE (right y-axis) w.r.t different $\beta$. The performance improves along with significant fairness improvement. \textbf{Left to Right:} ML-100K (G), ML-100K (A), ML-1M (G), ML-1M (A), Amazon-Movies(A)}
\label{fig:ValDisp&Change}
\end{figure*}
\subsubsection{Evaluation metric:} The performance of F$^{2}$PGNN for recommendations is measured by rooted mean square error (RMSE). Lower values of RMSE correspond to better recommendations. To quantify fairness, we use the difference principle as per Eq.(\ref{Eq:disp_def}). Lower values of $\mathcal{L}_{fair}$ indicate a fair model.

\subsection{Results} \label{sec:Results}
We compared the performance of F$^{2}$PGNN under different fairness budget levels against the baseline as described in Section \ref{sec:expsetup}. Table \ref{Table:TestResults} summarizes the results; the best results are shown in bold, and hyperparameter settings are given in Appendix \ref{Hyperparameter}. \par
F$^{2}$PGNN outperforms the baseline for all $\beta$ in terms of fairness across all datasets. At the same time, in terms of utility (RMSE), F$^{2}$PGNN outperforms all datasets except Amazon Movies. For F2MF, when the value of $\beta$ is larger than some threshold,  the model shows inconsistent and unstable behaviour (threshold values are shown in the box). Whereas the performance of F$^{2}$PGNN is consistent and stable for all values of $\beta$ except in ML-100K (A). \textbf{F$^{2}$PGNN improves the fairness in ML-100K (G), ML-100K (A), ML-1M (G), ML-1M (A) and Amazon Movies (A) by $\mathbf{47.65\%}$, $\mathbf{54.92\%}$, $\mathbf{84.22\%}$, $\mathbf{95.21\%}$ and $\mathbf{99.78\%}$, respectively, corresponding to threshold of $\mathbf{\beta}$. Also, the gain in utility is $\mathbf{1.12\% \sim 15.9\%}$ over all datasets}, at the expense of $10.58\%$ increase in RMSE for Amazon Movies (A). However, it should be noted that the inherent group disparity (i.e. $\beta = 0$) for F$^{2}$PGNN is less as compared to that of F2MF due to the fact that F2MF updates the embeddings based on explicit user-item interaction, whereas F$^{2}$PGNN considers higher-order interactions between the users to update the embeddings.

 \subsubsection{Performance Analysis for different fairness budgets $\left(\beta \right)$:} 
\begin{figure*}[!h]
\includegraphics[width=.195\textwidth]{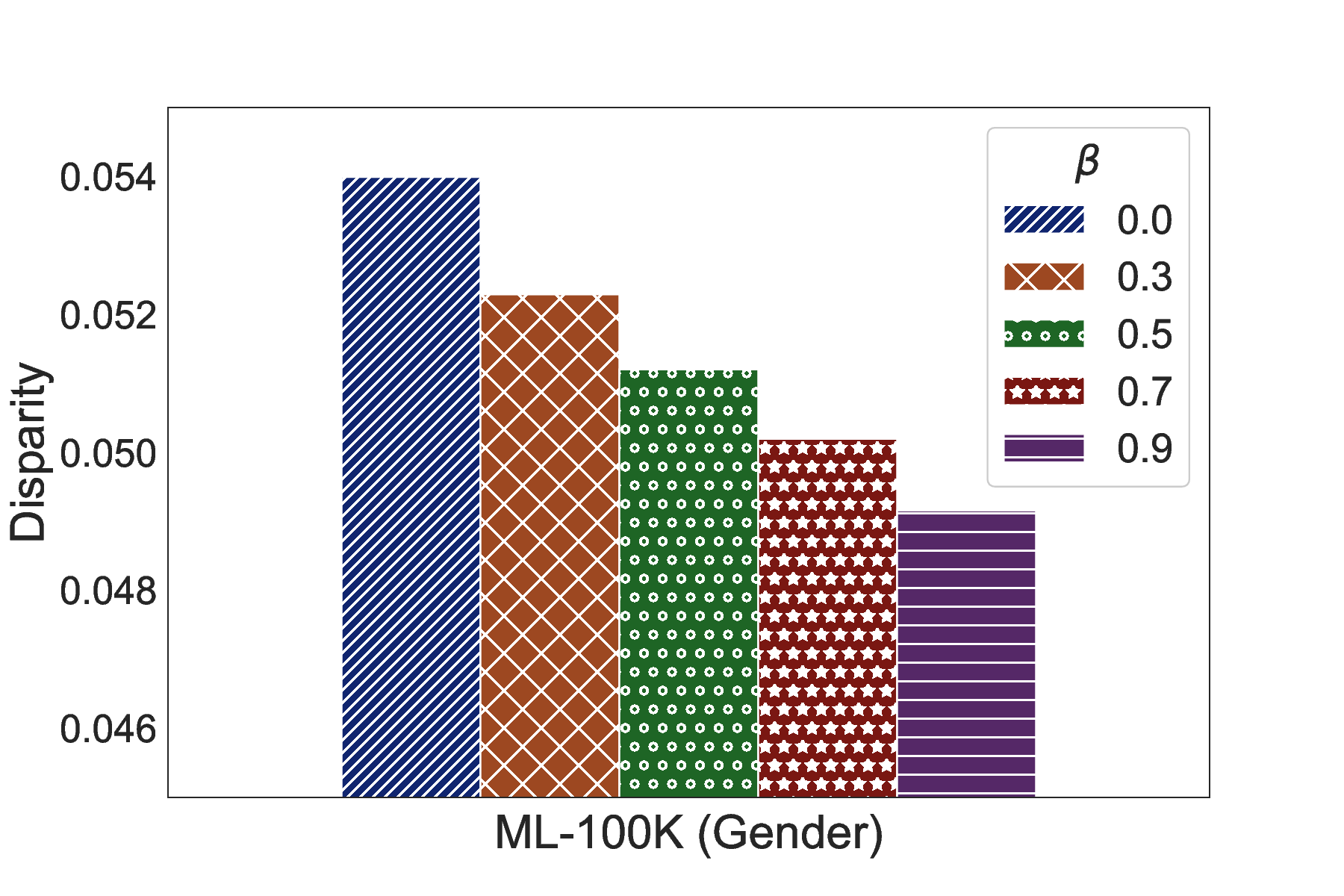}
\includegraphics[width=.195\textwidth]{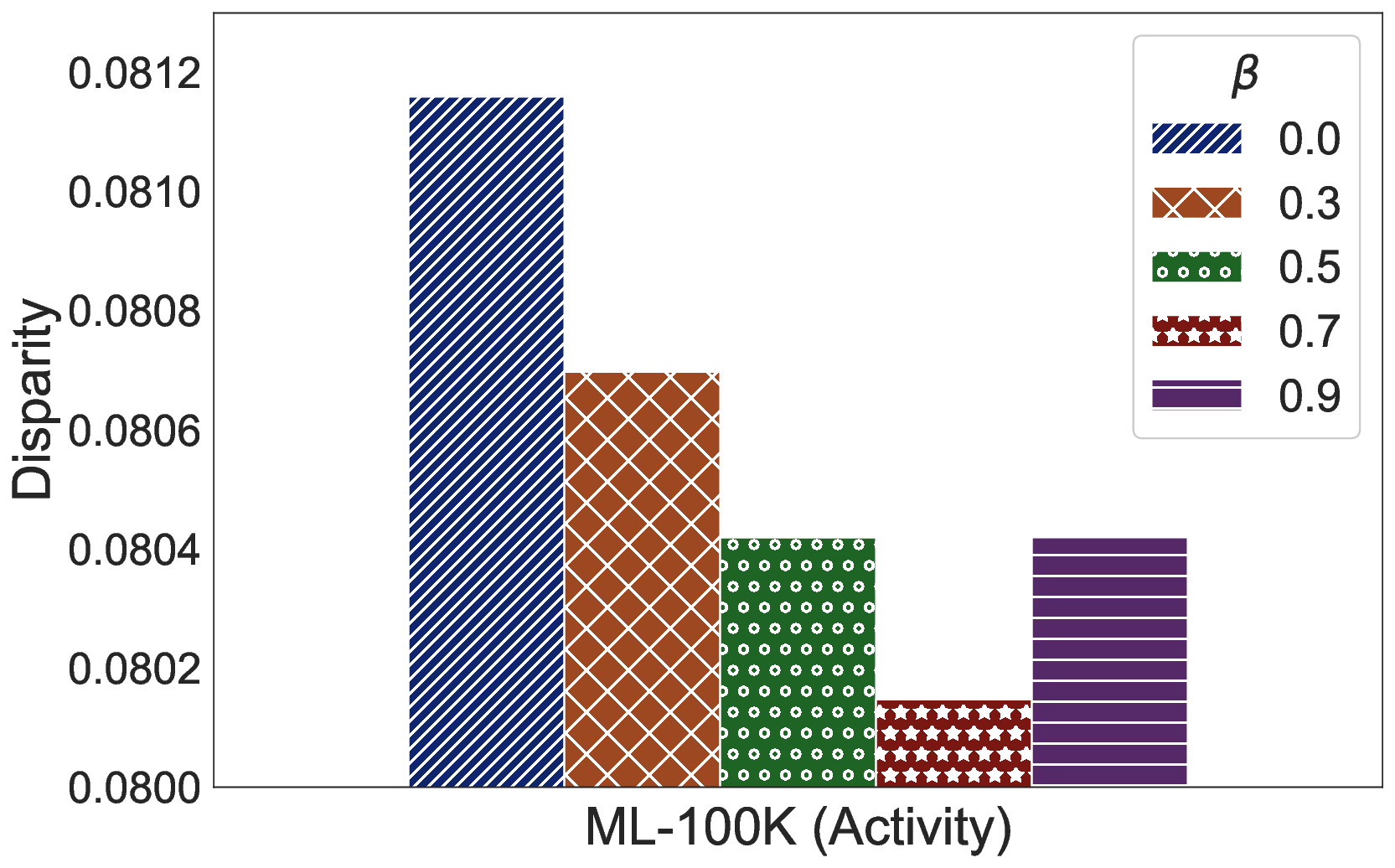} 
\includegraphics[width=.195\textwidth]{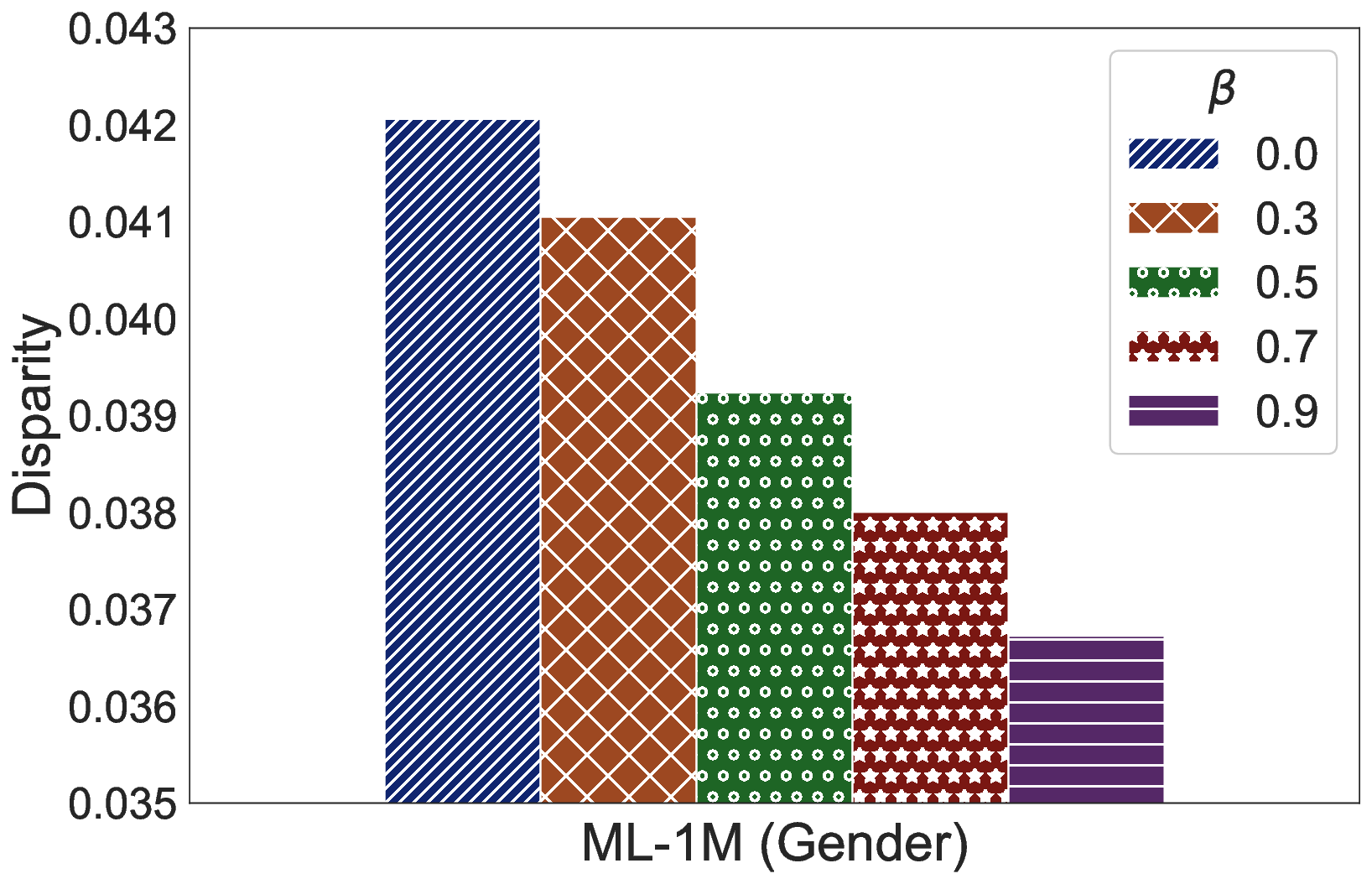}
\includegraphics[width=.195\textwidth]{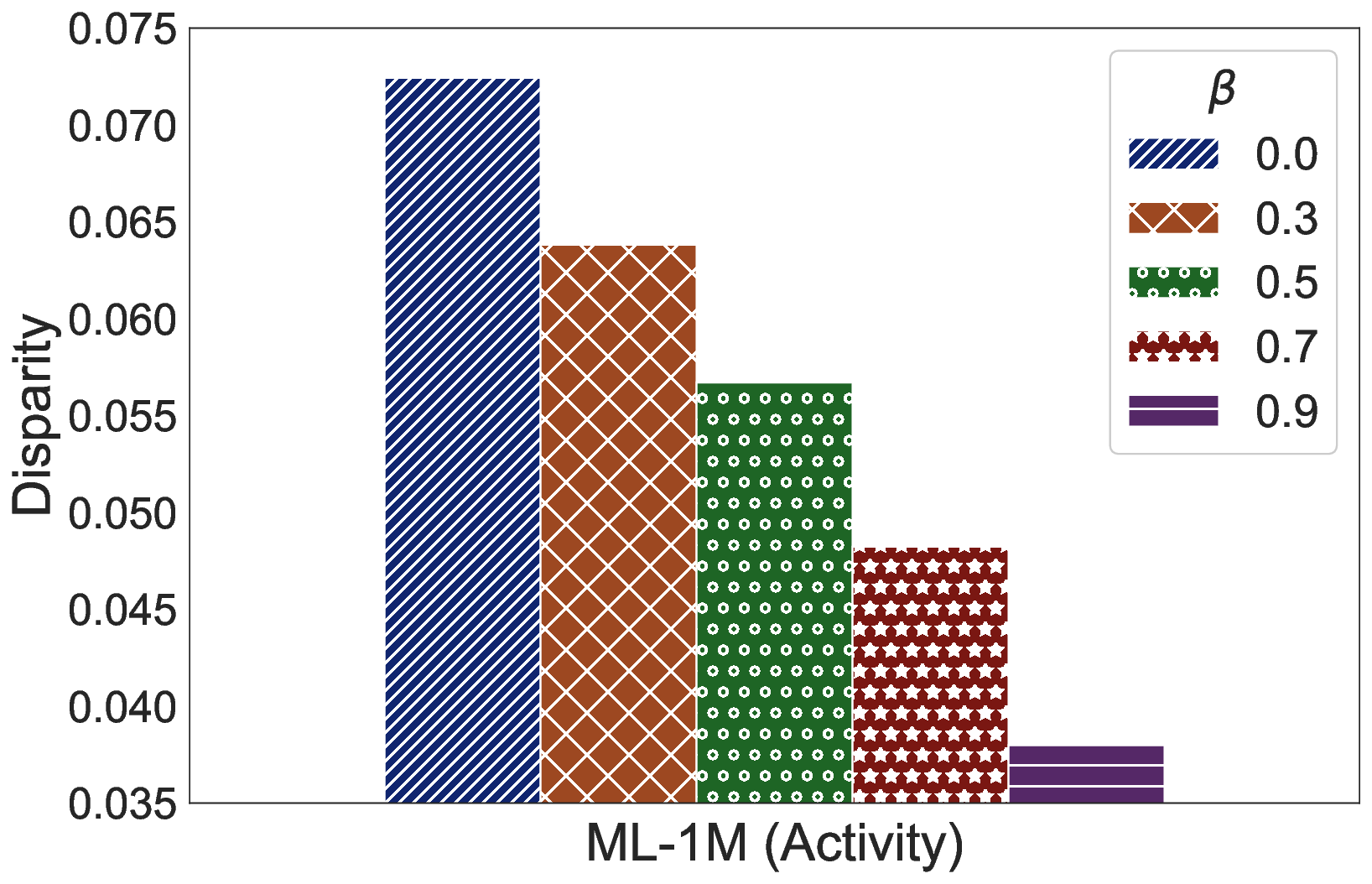}%
\includegraphics[width=.195\textwidth]{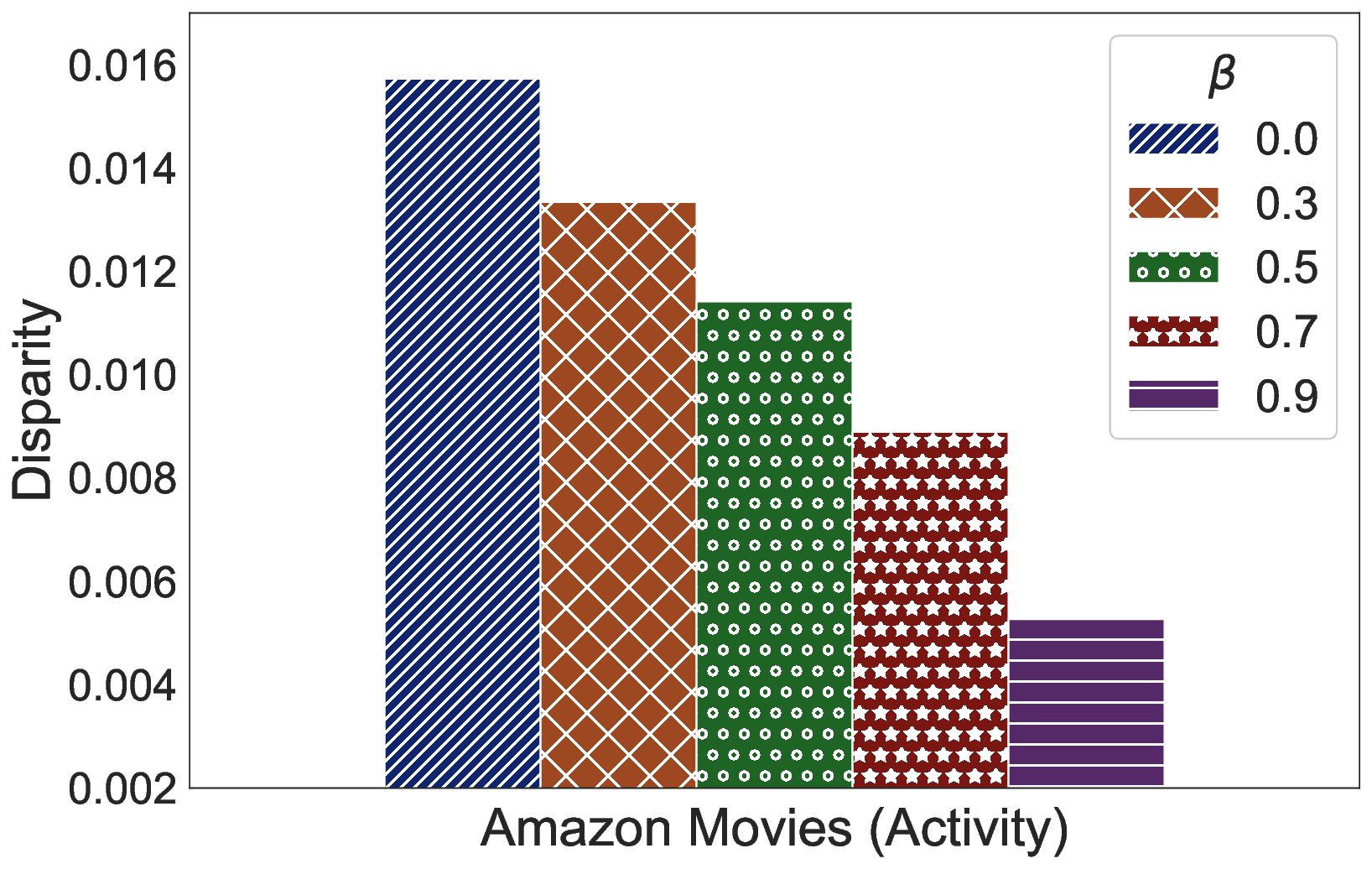}
\caption{Disparity on test data for different $\beta$. Increasing $\beta$ leads to reduced disparity.}
\label{fig:Testdisp}
\end{figure*}

For F$^{2}$PGNN, the parameter $\beta$ controls how much weightage is be given to the $\mathcal{L}_{fair}$ for fairness adaptation in each communication round. The top row of Figure \ref{fig:ValDisp&Change} visualizes the impact of $\beta$ on the group disparity for validation data. It is observed that the fairness constraint becomes prominent, yielding better fairness (curve goes down) as the value of $\beta$ increases. The bottom row of Figure \ref{fig:ValDisp&Change} showcases the fairness-utility trade-off. \textbf{The $\mathbf{\%}$ improvement in fairness is $\mathbf{9\% \sim 67\%}$ over all the datasets while maintaining the utility ( $\mathbf{\sim 1\% -2.5\%}$ reduction except $\mathbf{2.5\%}$ increase in RMSE for ML-100K(G))}. This is due to the fact that the model gets regularized better by incorporating fairness constraint for ML-1M and Amazon-Movies data whereas it is not significant for ML-100K. Similar to F2MF, for F$^{2}$PGNN, we observe unstable behavior for ML-100K (A) after a certain threshold of $\beta$. Figure \ref{fig:Testdisp} visualizes the trend in group disparity on the test data over all the datasets. When $\beta$ increases, the disparity consistently decreases except for ML-100K (A), which is evident from Figure \ref{fig:ValDisp&Change} (Bottom row). This shows that F$^{2}$PGNN is highly effective in achieving fairness while maintaining utility.

\begin{figure*}[!h]
\centering
\includegraphics[width=.81\textwidth]{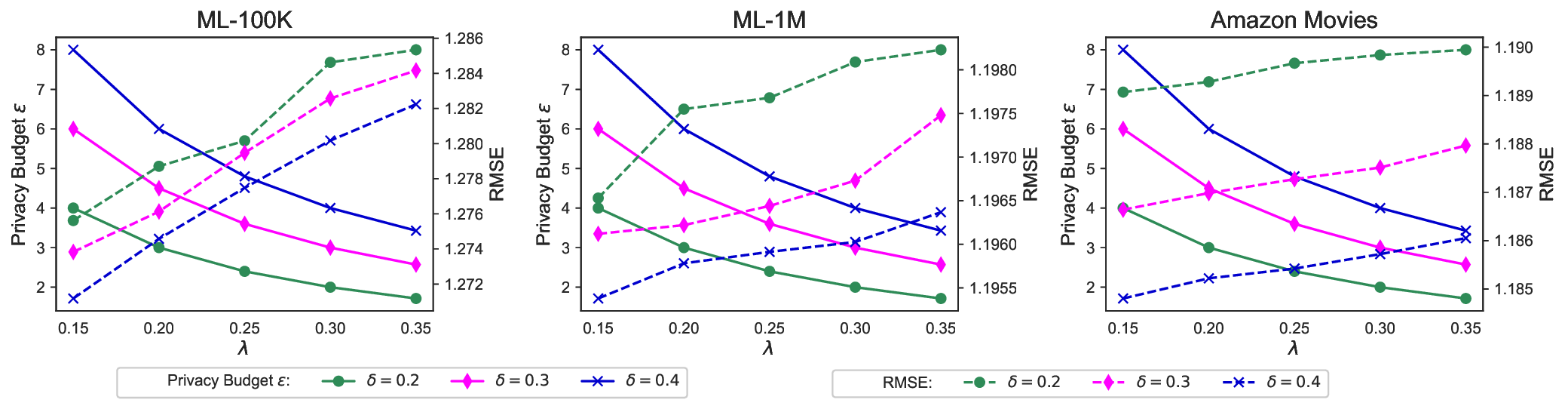}\quad
\caption{
Privacy budget $\epsilon$ 
(left y-axis) and the personalization RMSE (right y-axis) w.r.t different clipping threshold $\delta$ and noise variance $\lambda$. Lower $\epsilon$ implies better privacy and lower RMSE implies better performance. The performance sacrifice is higher for the lowest $\delta$ and it also goes higher with increasing $\lambda$. \textbf{Left to Right:} ML-100K (G), ML-1M (G), Amazon-Movies(A)}
\label{fig:LDPAnalysis}
\end{figure*}
 \subsection{Ablation Study} \label{sec:Ablation}
To get better understanding of how different hyperparameters influence different aspects of F$^{2}$PGNN, including performance and privacy protection, we conduct ablation studies to analyse the effect of these parameters.
\subsubsection{Performance Analysis with LDP:}
There might be privacy issues if the GNN model parameters and item embedding gradients are uploaded directly to the server \cite{Zhu2019}. The reason being only the item that user has interacted will have non-zero gradients, hence server can easily identify the user interaction history. F2MF considered LDP only in the group statistics which is not sufficient for user privacy protection. To overcome this caveat, we consider LDP for ensuring privacy as described in Section \ref{sec:PrivacyProtection}.\par
We first analyze the privacy-utility trade-off, by varying $\delta$ and $\lambda$ in the LDP module (Figure \ref{fig:LDPAnalysis}). Smaller values of $\delta$ and larger values of $\lambda$ incur a smaller privacy budget $\epsilon$, which implies better privacy protection. Note that, the fairness budget $\beta$ is set to $0.5$ for analysis. The model performance of F$^{2}$PGNN declines with the increase in noise strength $\lambda$. Moreover, smaller gradient clipping thresholds such as $\delta = 0.2$ deteriorates the prediction substantially. \textbf{If observed carefully, the worst increase in RMSE w.r.t one without gradient clipping is $\mathbf{1.88\%}$, $\mathbf{0.27\%}$ and $\mathbf{0.474\%}$ for ML-100K (G), ML-1M (G) and Amazon Movies (A) respectively}. Thus, selecting $\delta$ and $\lambda$ to strike a balance between privacy protection and recommendation accuracy is critical. We further investigate the effect of LDP on privacy-fairness tradeoff, deferred to Appendix \ref{additional_ldp} in the interest of space.
\subsubsection{LDP on Group Statistics:}
\begin{figure}[h]
\centering 
\includegraphics[width=.23\textwidth]{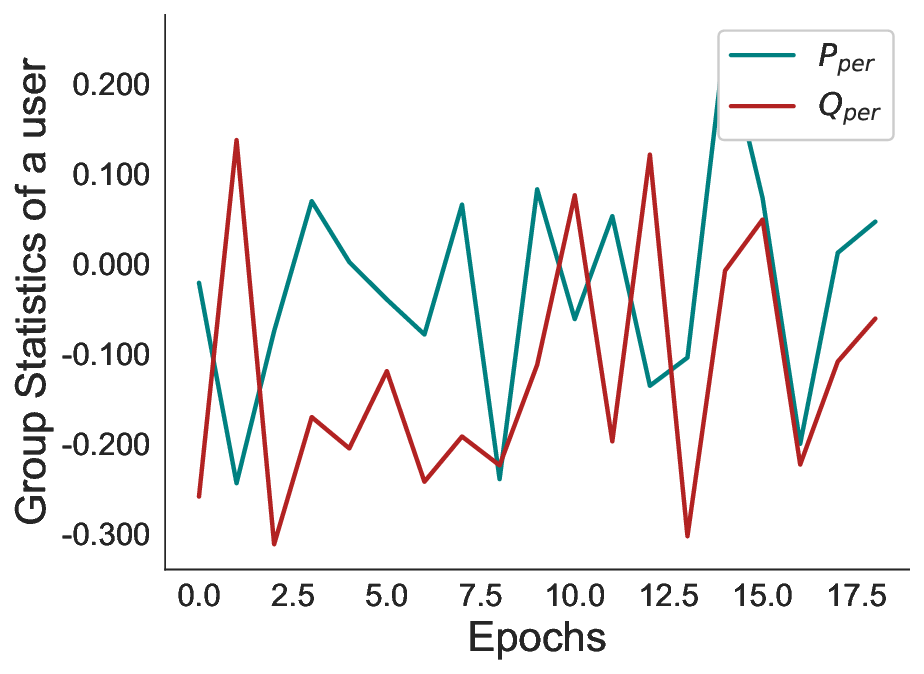}
\includegraphics[width=.23\textwidth]{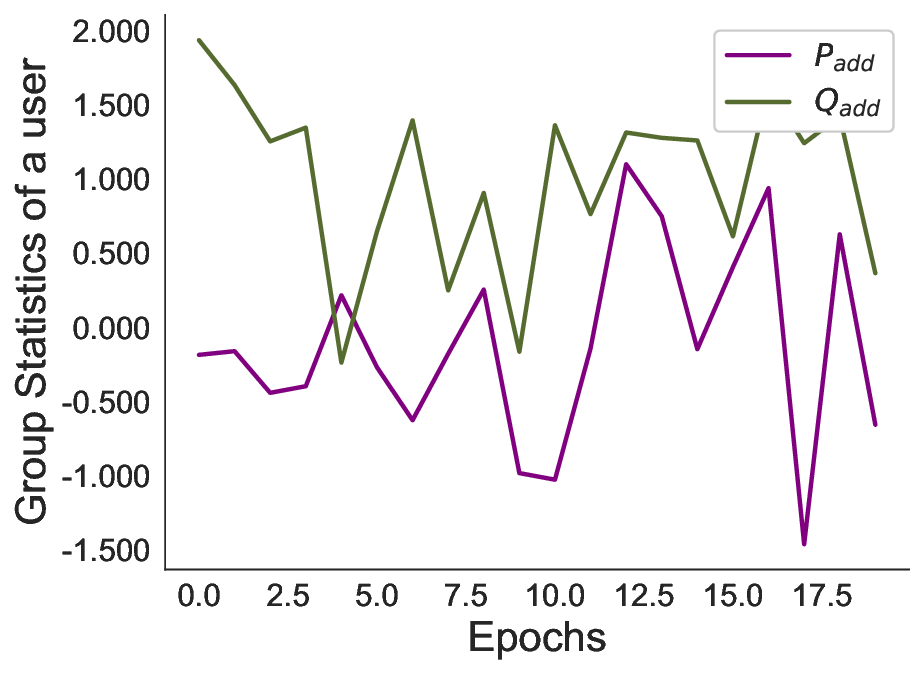}%
\caption{Effect of LDP on group statistics for a given user of group $S_0$ for fixed $\beta$, $\delta$, $\lambda$ and $\sigma$ combination for Amazon Movies (Activity). }
\label{fig:GroupStatLDP}
\end{figure}

The update of group statistics, i.e. $P$ and $Q$ requires demographic information from each user without exposing their original identity. Hence, each users broadcasts the group statistics with LDP (refer Section \ref{sec:PrivacyProtection}), to avoid revealing the true group membership. Figure \ref{fig:GroupStatLDP} illustrates how the local statistics for a particular user hides the users' true demographic features through LDP. Ideally, the value of $P_{per}^{u} \ \& \ Q_{per}^{u}$ must be $\frac{- \mathcal{L}_{util}^{u}}{2}$ and $P_{add}^{u} \ \& \ Q_{add}^{u}$ be $0.5$ making it impossible for server to infer user attribute. But this comes at the expense of losing fairness and utility of the model as the strength of noise required is significantly high. As the group statistics are intermingled, the server cannot infer the users' group membership, as shown in the analysis for Amazon Movies (A) (Figure \ref{fig:GroupStatLDP}), for other datasets results are presented in the Appendix \ref{ldp_gstat}.

\vspace{-0.7cm}
\section{Conclusion} \label{sec:Conclusion}
In this work, motivated by the importance of GNN in FRS, and the challenges related to inherent bias in the model, we present F$^{2}$PGNN, a novel framework for a privacy-preserved global group fair recommendation system. We introduce a privacy-preserved inductive graph expansion technique that minimizes communication overhead. We also enhance privacy protection through an LDP module while broadcasting the model and group statistics update to the server. For F$^{2}$PGNN, we empirically demonstrate improvements (and associated trade-offs) over the state-of-the-art in terms of accuracy, efficiency and fairness. F$^{2}$PGNN does not consider asynchronous updates, thus it can be an interesting future direction to develop a GNN-based fair FRS by incorporating computational heterogeneity. 
\bibliography{aaai24}

\begin{thebibliography}{39}
\providecommand{\natexlab}[1]{#1}

\bibitem[{Aledhari et~al.(2020)Aledhari, Razzak, Parizi, and
  Saeed}]{aledhari2020}
Aledhari, M.; Razzak, R.; Parizi, R.~M.; and Saeed, F. 2020.
\newblock Federated learning: A survey on enabling technologies, protocols, and
  applications.
\newblock \emph{IEEE Access}, 8: 140699--140725.

\bibitem[{Ammad-Ud-Din et~al.(2019)Ammad-Ud-Din, Ivannikova, Khan, Oyomno, Fu,
  Tan, and Flanagan}]{ammad2019federated}
Ammad-Ud-Din, M.; Ivannikova, E.; Khan, S.~A.; Oyomno, W.; Fu, Q.; Tan, K.~E.;
  and Flanagan, A. 2019.
\newblock Federated collaborative filtering for privacy-preserving personalized
  recommendation system.
\newblock \emph{arXiv preprint arXiv:1901.09888}.

\bibitem[{Berg, Kipf, and Welling(2017)}]{berg2017graph}
Berg, R. v.~d.; Kipf, T.~N.; and Welling, M. 2017.
\newblock Graph convolutional matrix completion.
\newblock \emph{arXiv preprint arXiv:1706.02263}.

\bibitem[{Chai et~al.(2020)Chai, Wang, Chen, and Yang}]{chai2020secure}
Chai, D.; Wang, L.; Chen, K.; and Yang, Q. 2020.
\newblock Secure federated matrix factorization.
\newblock \emph{IEEE Intelligent Systems}, 36(5): 11--20.

\bibitem[{Choi et~al.(2018)Choi, Tomei, Vicarte, Hanumolu, and
  Kumar}]{choi2018guaranteeing}
Choi, W.-S.; Tomei, M.; Vicarte, J. R.~S.; Hanumolu, P.~K.; and Kumar, R. 2018.
\newblock Guaranteeing local differential privacy on ultra-low-power systems.
\newblock In \emph{2018 ACM/IEEE 45th Annual International Symposium on
  Computer Architecture (ISCA)}, 561--574.

\bibitem[{Du et~al.(2021)Du, Xu, Wu, and Tong}]{du2021fairness}
Du, W.; Xu, D.; Wu, X.; and Tong, H. 2021.
\newblock Fairness-aware agnostic federated learning.
\newblock In \emph{Proceedings of the 2021 SIAM International Conference on
  Data Mining (SDM)}, 181--189.

\bibitem[{Dwork et~al.(2012)Dwork, Hardt, Pitassi, Reingold, and
  Zemel}]{Dwork2012}
Dwork, C.; Hardt, M.; Pitassi, T.; Reingold, O.; and Zemel, R. 2012.
\newblock Fairness through Awareness.
\newblock In \emph{Proceedings of the 3rd Innovations in Theoretical Computer
  Science Conference}, 214–226. New York, NY, USA: Association for Computing
  Machinery.

\bibitem[{Dwork and Roth(2014)}]{Dwork14}
Dwork, C.; and Roth, A. 2014.
\newblock The Algorithmic Foundations of Differential Privacy.
\newblock 9(3–4): 211–407.

\bibitem[{Gao, Ge, and Shah(2022)}]{gao2022fair}
Gao, R.; Ge, Y.; and Shah, C. 2022.
\newblock FAIR: Fairness-aware information retrieval evaluation.
\newblock \emph{Journal of the Association for Information Science and
  Technology}, 73(10): 1461--1473.

\bibitem[{Hamilton, Ying, and Leskovec(2017)}]{graphsage}
Hamilton, W.; Ying, Z.; and Leskovec, J. 2017.
\newblock Inductive Representation Learning on Large Graphs.
\newblock \emph{Advances in Neural Information Processing Systems}, 30.

\bibitem[{Hardt et~al.(2016)Hardt, Price, Price, and Srebro}]{Hardt2016}
Hardt, M.; Price, E.; Price, E.; and Srebro, N. 2016.
\newblock Equality of Opportunity in Supervised Learning.
\newblock \emph{Advances in Neural Information Processing Systems}, 29.

\bibitem[{Harper and Konstan(2015)}]{Harper15}
Harper, F.~M.; and Konstan, J.~A. 2015.
\newblock The MovieLens Datasets: History and Context.
\newblock \emph{ACM Transactions on Interactive Intelligent Systems}, 5(4):
  1–19.

\bibitem[{He et~al.(2021)He, Balasubramanian, Ceyani, Yang, Xie, Sun, He, Yang,
  Yu, Rong et~al.}]{he2021fedgraphnn}
He, C.; Balasubramanian, K.; Ceyani, E.; Yang, C.; Xie, H.; Sun, L.; He, L.;
  Yang, L.; Yu, P.~S.; Rong, Y.; et~al. 2021.
\newblock Fedgraphnn: A federated learning system and benchmark for graph
  neural networks.
\newblock \emph{arXiv preprint arXiv:2104.07145}.

\bibitem[{He et~al.(2020)He, Deng, Wang, Li, Zhang, and Wang}]{he2020lightgcn}
He, X.; Deng, K.; Wang, X.; Li, Y.; Zhang, Y.; and Wang, M. 2020.
\newblock Lightgcn: Simplifying and powering graph convolution network for
  recommendation.
\newblock In \emph{Proceedings of the 43rd International ACM SIGIR conference
  on research and development in Information Retrieval}, 639--648.

\bibitem[{Islam et~al.(2019)Islam, Keya, Pan, and Foulds}]{islam2019mitigating}
Islam, R.; Keya, K.~N.; Pan, S.; and Foulds, J. 2019.
\newblock Mitigating demographic biases in social media-based recommender
  systems.
\newblock \emph{KDD (Social Impact Track)}.

\bibitem[{Kairouz et~al.(2021)Kairouz, McMahan, Avent, Bellet, Bennis, Bhagoji,
  Bonawitz, Charles, Cormode, Cummings et~al.}]{kairouz2021}
Kairouz, P.; McMahan, H.~B.; Avent, B.; Bellet, A.; Bennis, M.; Bhagoji, A.~N.;
  Bonawitz, K.; Charles, Z.; Cormode, G.; Cummings, R.; et~al. 2021.
\newblock Advances and open problems in federated learning.
\newblock \emph{Foundations and Trends{\textregistered} in Machine Learning},
  14(1--2): 1--210.

\bibitem[{Kipf and Welling(2017)}]{kipf2017}
Kipf, T.~N.; and Welling, M. 2017.
\newblock Semi-Supervised Classification with Graph Convolutional Networks.
\newblock In \emph{International Conference on Learning Representations}.

\bibitem[{Koren, Bell, and Volinsky(2009)}]{koren2009matrix}
Koren, Y.; Bell, R.; and Volinsky, C. 2009.
\newblock Matrix factorization techniques for recommender systems.
\newblock \emph{Computer}, 42(8): 30--37.

\bibitem[{Li et~al.(2021{\natexlab{a}})Li, Chen, Fu, Ge, and
  Zhang}]{li2021user}
Li, Y.; Chen, H.; Fu, Z.; Ge, Y.; and Zhang, Y. 2021{\natexlab{a}}.
\newblock User-oriented fairness in recommendation.
\newblock In \emph{Proceedings of the Web Conference 2021}, 624--632.

\bibitem[{Li et~al.(2021{\natexlab{b}})Li, Chen, Fu, Ge, and Zhang}]{Yunqi2021}
Li, Y.; Chen, H.; Fu, Z.; Ge, Y.; and Zhang, Y. 2021{\natexlab{b}}.
\newblock User-Oriented Fairness in Recommendation.
\newblock In \emph{Proceedings of the Web Conference 2021}, WWW '21, 624–632.
  New York, NY, USA: Association for Computing Machinery.
\newblock ISBN 9781450383127.

\bibitem[{Li, Ge, and Zhang(2021)}]{li2021tutorial}
Li, Y.; Ge, Y.; and Zhang, Y. 2021.
\newblock Tutorial on fairness of machine learning in recommender systems.
\newblock In \emph{Proceedings of the 44th international ACM SIGIR conference
  on research and development in information retrieval}, 2654--2657.

\bibitem[{Liu et~al.(2022{\natexlab{a}})Liu, Ge, Xu, Zhang, and
  Marian}]{Liu2022}
Liu, S.; Ge, Y.; Xu, S.; Zhang, Y.; and Marian, A. 2022{\natexlab{a}}.
\newblock Fairness-Aware Federated Matrix Factorization.
\newblock In \emph{Proceedings of the 16th ACM Conference on Recommender
  Systems}, RecSys '22, 168–178. New York, NY, USA: Association for Computing
  Machinery.

\bibitem[{Liu et~al.(2022{\natexlab{b}})Liu, Yang, Fan, Peng, and
  Yu}]{liu2022federated}
Liu, Z.; Yang, L.; Fan, Z.; Peng, H.; and Yu, P.~S. 2022{\natexlab{b}}.
\newblock Federated social recommendation with graph neural network.
\newblock \emph{ACM Transactions on Intelligent Systems and Technology (TIST)},
  13(4): 1--24.

\bibitem[{Maeng et~al.(2022)Maeng, Lu, Melis, Nguyen, Rabbat, and
  Wu}]{maeng2022towards}
Maeng, K.; Lu, H.; Melis, L.; Nguyen, J.; Rabbat, M.; and Wu, C.-J. 2022.
\newblock Towards fair federated recommendation learning: Characterizing the
  inter-dependence of system and data heterogeneity.
\newblock In \emph{Proceedings of the 16th ACM Conference on Recommender
  Systems}, 156--167.

\bibitem[{Magdziarczyk(2019)}]{magdziarczyk2019right}
Magdziarczyk, M. 2019.
\newblock Right to be forgotten in light of regulation (eu) 2016/679 of the
  european parliament and of the council of 27 april 2016 on the protection of
  natural persons with regard to the processing of personal data and on the
  free movement of such data, and repealing directive 95/46/ec.
\newblock In \emph{6th International Multidisciplinary Scientific Conference on
  Social Sciences and Art Sgem 2019}, 177--184.

\bibitem[{McMahan et~al.(2017)McMahan, Moore, Ramage, Hampson, and
  y~Arcas}]{mcmahan2017}
McMahan, B.; Moore, E.; Ramage, D.; Hampson, S.; and y~Arcas, B.~A. 2017.
\newblock Communication-efficient learning of deep networks from decentralized
  data.
\newblock In \emph{Artificial intelligence and statistics}, 1273--1282. PMLR.

\bibitem[{McSherry and Mironov(2009)}]{mcsherry2009differentially}
McSherry, F.; and Mironov, I. 2009.
\newblock Differentially private recommender systems: Building privacy into the
  netflix prize contenders.
\newblock In \emph{Proceedings of the 15th ACM SIGKDD international conference
  on Knowledge discovery and data mining}, 627--636.

\bibitem[{Ni, Li, and McAuley(2019)}]{ni-etal-2019}
Ni, J.; Li, J.; and McAuley, J. 2019.
\newblock Justifying Recommendations using Distantly-Labeled Reviews and
  Fine-Grained Aspects.
\newblock In \emph{Proceedings of the 2019 Conference on Empirical Methods in
  Natural Language Processing and the 9th International Joint Conference on
  Natural Language Processing (EMNLP-IJCNLP)}, 188--197. Hong Kong, China:
  Association for Computational Linguistics.

\bibitem[{Qi et~al.(2020)Qi, Wu, Wu, Huang, and Xie}]{qi-etal-2020}
Qi, T.; Wu, F.; Wu, C.; Huang, Y.; and Xie, X. 2020.
\newblock Privacy-Preserving News Recommendation Model Learning.
\newblock In \emph{Findings of the Association for Computational Linguistics:
  EMNLP 2020}, 1423--1432. Online: Association for Computational Linguistics.

\bibitem[{Sarwar et~al.(2000)Sarwar, Karypis, Konstan, and Riedl}]{sarwar2000}
Sarwar, B.; Karypis, G.; Konstan, J.; and Riedl, J. 2000.
\newblock Analysis of recommendation algorithms for e-commerce.
\newblock In \emph{Proceedings of the 2nd ACM Conference on Electronic
  Commerce}, 158--167.

\bibitem[{Veličković et~al.(2018)Veličković, Cucurull, Casanova, Romero,
  Liò, and Bengio}]{gat2018}
Veličković, P.; Cucurull, G.; Casanova, A.; Romero, A.; Liò, P.; and Bengio,
  Y. 2018.
\newblock Graph Attention Networks.
\newblock In \emph{International Conference on Learning Representations}.

\bibitem[{Wang et~al.(2019)Wang, He, Wang, Feng, and Chua}]{wang2019neural}
Wang, X.; He, X.; Wang, M.; Feng, F.; and Chua, T.-S. 2019.
\newblock Neural graph collaborative filtering.
\newblock In \emph{Proceedings of the 42nd international ACM SIGIR conference
  on Research and development in Information Retrieval}, 165--174.

\bibitem[{Wang et~al.(2023)Wang, Ma, Zhang, Liu, and Ma}]{Wang2023}
Wang, Y.; Ma, W.; Zhang, M.; Liu, Y.; and Ma, S. 2023.
\newblock A Survey on the Fairness of Recommender Systems.
\newblock \emph{ACM Transactions on Information Systems}, 41(3): 1–43.

\bibitem[{Wang et~al.(2021)Wang, Fan, Qi, Wen, Wang, and
  Yu}]{wang2021federated}
Wang, Z.; Fan, X.; Qi, J.; Wen, C.; Wang, C.; and Yu, R. 2021.
\newblock Federated Learning with Fair Averaging.
\newblock In Zhou, Z.-H., ed., \emph{Proceedings of the Thirtieth International
  Joint Conference on Artificial Intelligence, {IJCAI-21}}, 1615--1623.
  International Joint Conferences on Artificial Intelligence Organization.

\bibitem[{Wu et~al.(2022{\natexlab{a}})Wu, Wu, Lyu, Qi, Huang, and
  Xie}]{wu2022federated}
Wu, C.; Wu, F.; Lyu, L.; Qi, T.; Huang, Y.; and Xie, X. 2022{\natexlab{a}}.
\newblock A federated graph neural network framework for privacy-preserving
  personalization.
\newblock \emph{Nature Communications}, 13(1): 3091.

\bibitem[{Wu et~al.(2022{\natexlab{b}})Wu, Sun, Zhang, Xie, and
  Cui}]{wu2022graph}
Wu, S.; Sun, F.; Zhang, W.; Xie, X.; and Cui, B. 2022{\natexlab{b}}.
\newblock Graph neural networks in recommender systems: a survey.
\newblock \emph{ACM Computing Surveys}, 55(5): 1--37.

\bibitem[{Yao and Huang(2017)}]{yao2017beyond}
Yao, S.; and Huang, B. 2017.
\newblock Beyond parity: Fairness objectives for collaborative filtering.
\newblock \emph{Advances in neural information processing systems}, 30.

\bibitem[{Ying et~al.(2018)Ying, He, Chen, Eksombatchai, Hamilton, and
  Leskovec}]{ying2018graph}
Ying, R.; He, R.; Chen, K.; Eksombatchai, P.; Hamilton, W.~L.; and Leskovec, J.
  2018.
\newblock Graph convolutional neural networks for web-scale recommender
  systems.
\newblock In \emph{Proceedings of the 24th ACM SIGKDD international conference
  on knowledge discovery \& data mining}, 974--983.

\bibitem[{Zhu, Liu, and Han(2019)}]{Zhu2019}
Zhu, L.; Liu, Z.; and Han, S. 2019.
\newblock Deep Leakage from Gradients.
\newblock \emph{Advances in neural information processing systems}, 32.

\end{thebibliography}
\clearpage
\appendix
\section*{Appendix}
\section{Schematic Diagram for Fairness Controller} \label{FC}
In order to achieve global group fairness at the central server, each user observes the group statistics (i.e $P$ and $Q$) received from the server and regulates the learning rate accordingly. If a user finds that it belongs to superior performing group, then it scales down the learning rate with factor $L$ else it scales up according to Eq.(\ref{eq:scale}). Before aggregation, each user employs LDP (as per Eq.(\ref{eq:clip_noise})) on the scaled gradients and performs the local updates to enforce the effect of scaling in model parameters as well as user/item embeddings. Also to update $P$ and $Q$ for the next communication round, each user updates the local group statistics as per Eq.(\ref{eq:g_stats}). Then user broadcasts these updates to the central server for aggregation. Next, the server aggregates this information to update the model parameters, user/item embeddings, and group statistics which are subsequently broadcasted back to users. This iterative process continues until convergence is achieved. 
\begin{figure}[!h]
    \centering
    \includegraphics[width = 3in]{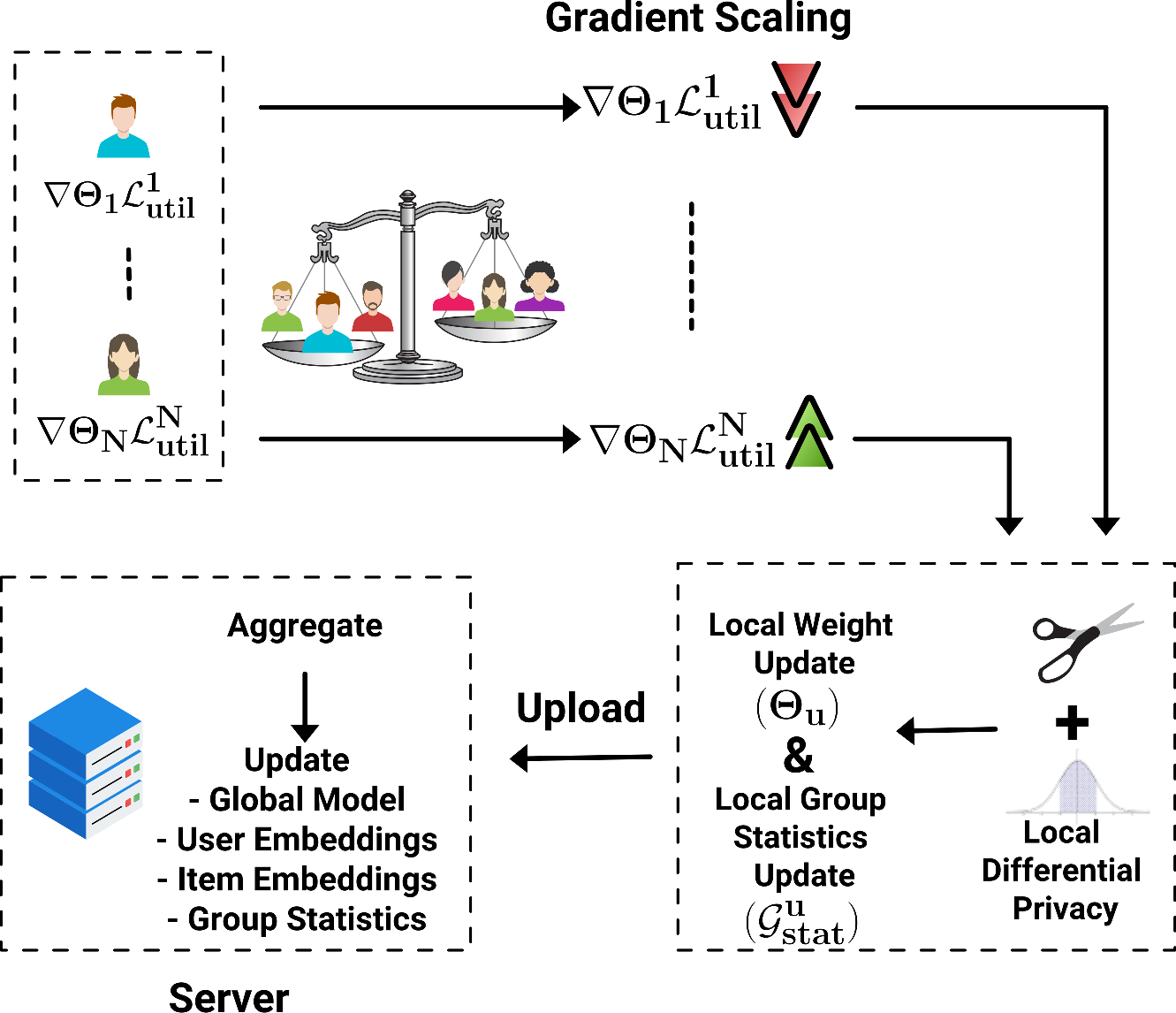}
    \caption{Framework of Fairness Controller for {\scshape F$^{2}$pgnn}}
    \label{fig:Fairnesscontroller}
\end{figure}
\section{Algorithm Pseudocode} \label{IG_expension}
In this section, we present the pseudocode for Privacy-Preserving Inductive Graph Expansion (Algorithm \ref{alg:privategraphalgorithm}) and Local Update (Algorithm \ref{alg:localupdate}).     
\begin{algorithm}[!h]
\caption{\textbf{Inductive Graph Expansion}}
\label{alg:privategraphalgorithm}
\textbf{Parameter}: $H$ - List of Item IDs interacted by user
\textbf{PrivateGraphExpansion$\left(\right)$ :}
\begin{algorithmic}[1] 
\STATE Send Public Key $E$ from the server to users
\FOR{each user $u \in U_K$}  
\STATE Update $H_{i}$ $\leftarrow$ Item IDs rated by $u$
\FOR{each item in $H_{i}$}  
\IF{item \textbf{in} $H_{i}$ \textbf{AND} item \textbf{not in} $H_{i-1}$}
\STATE Encrypt item with E
\ENDIF
\ENDFOR
\STATE Upload Encrypted IDs \& user embeddings to the server
\ENDFOR
\STATE Server forms a dictionary which maps each item with its neighbor embeddings
\RETURN Mapping-Dict
\end{algorithmic}
\end{algorithm}

\begin{algorithm}[!h]
\caption{\textbf{Local Update Function}}
\label{alg:localupdate}
\textbf{Parameter}: Clip norm $\left(\delta\right)$, Noise Parameter $\left(\lambda\right)$ \\
\textbf{Local Update$\left(\right)$ :}
\begin{algorithmic}[1] 
\STATE $\nabla_{\Theta}\mathcal{L}_{util}^{u}$ $\leftarrow$ $Clip\left ( \nabla_{\Theta}\mathcal{L}_{util}^{u}, \delta \right )$ $+$ $Laplace\left(0, \lambda\right)$
\STATE $\Theta_{u}^{i}$ $\leftarrow$ $\Theta_{u}^{i-1}$ $-$ $\eta \cdot \nabla_{\Theta}\mathcal{L}_{util}^{u} $
\RETURN $\Theta_{u}^{i}$ 
\end{algorithmic}
\end{algorithm}

\begin{figure*}[h]
\centering
\includegraphics[width=.3\textwidth]{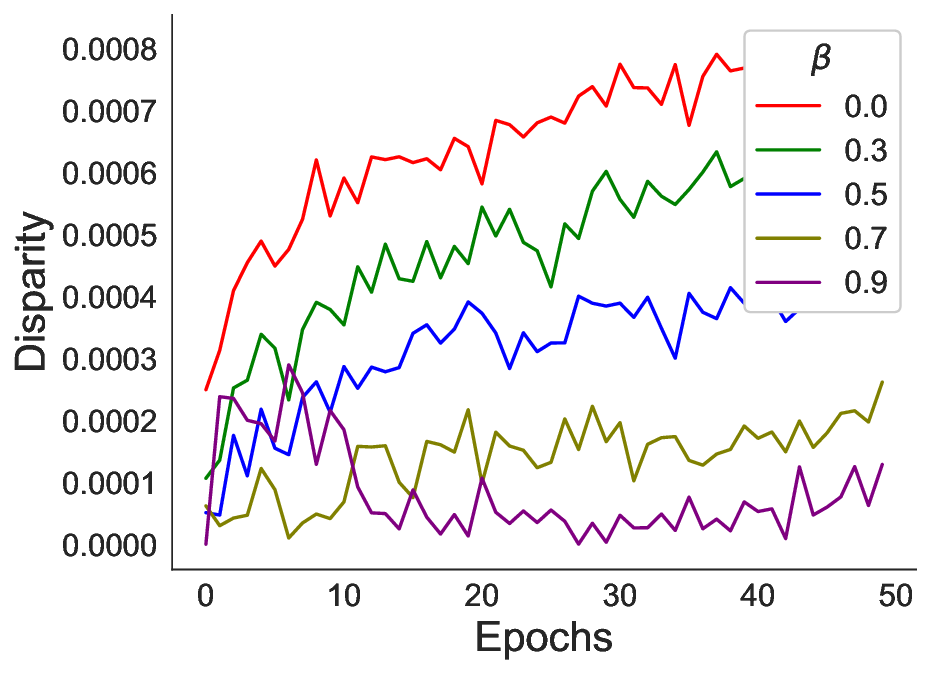}\quad  
\includegraphics[width=.3\textwidth]{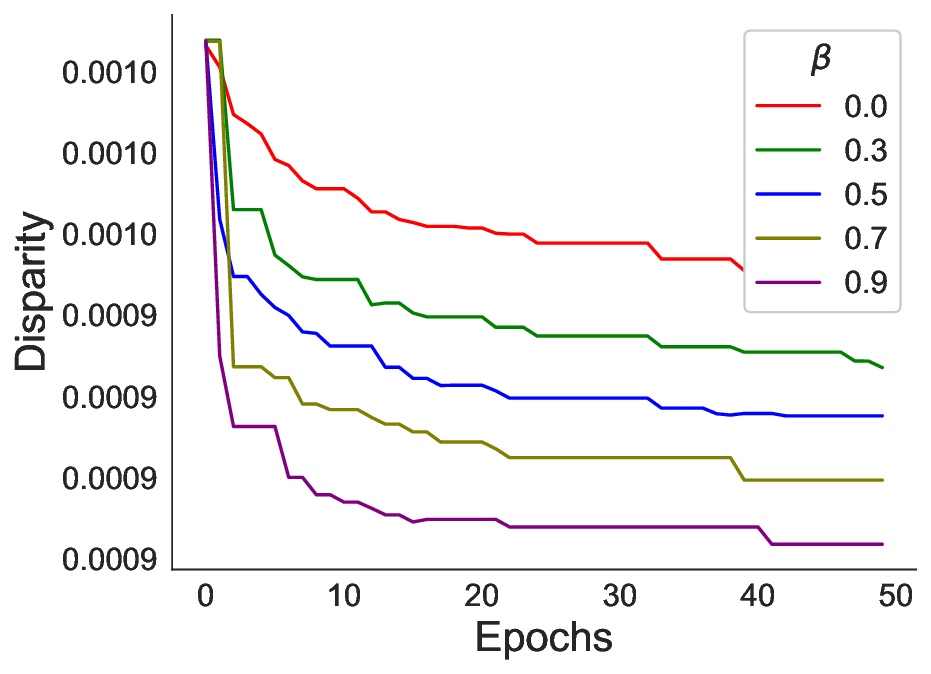}\quad    
\includegraphics[width=.3\textwidth]{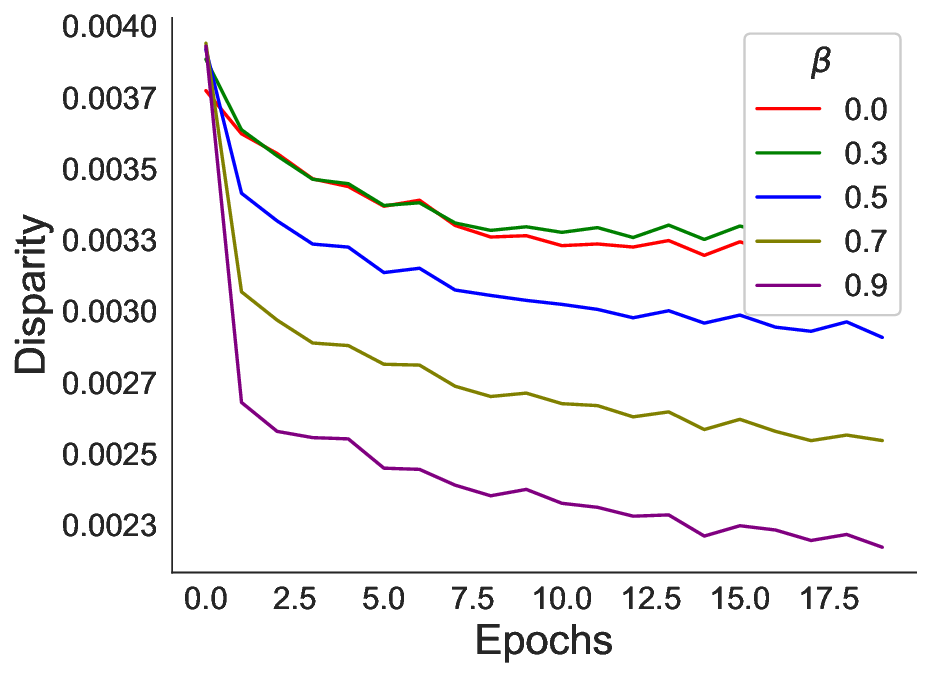}
\caption{
Disparity vs epochs with gradient clipping w.r.t different $\beta$ for fixed $\delta$ and $\lambda$. The curve gets noisy but still it goes down with increasing $\beta$. \textbf{Left to Right:} ML-100K (G), ML-1M (G), Amazon-Movies(A)}
\label{fig:LDPAnalysis2}
\end{figure*}
\begin{figure*}[h]
\centering 
\includegraphics[width=.23\textwidth]{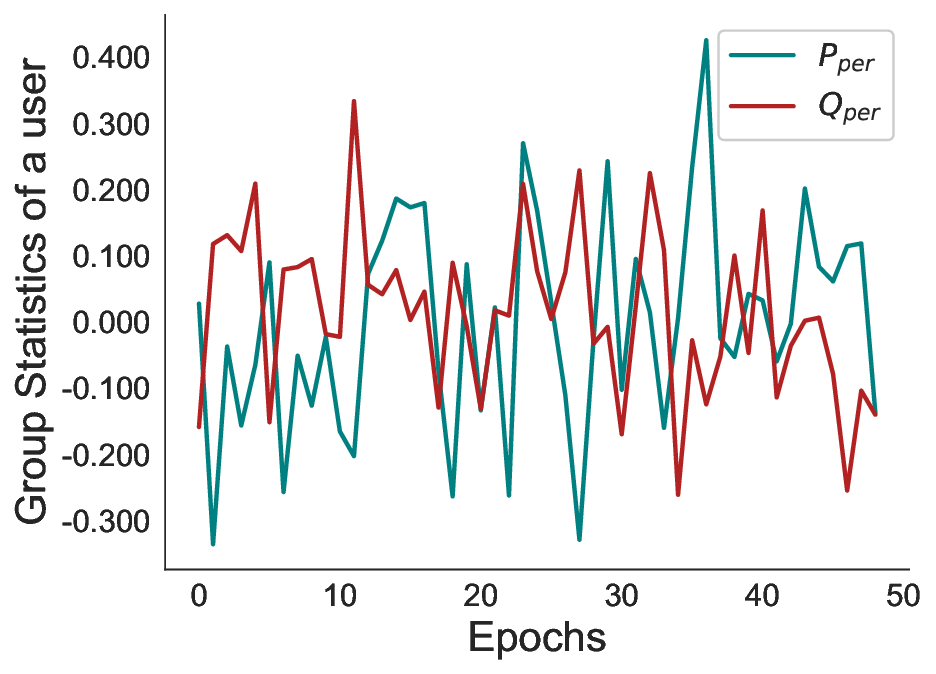}
\includegraphics[width=.23\textwidth]{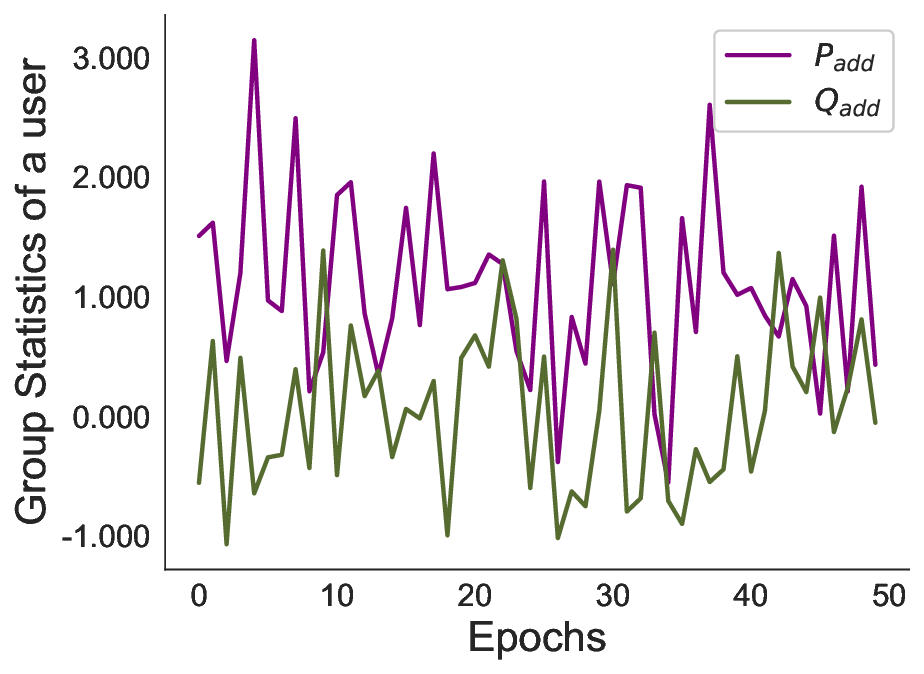}\quad
\includegraphics[width=.23\textwidth]{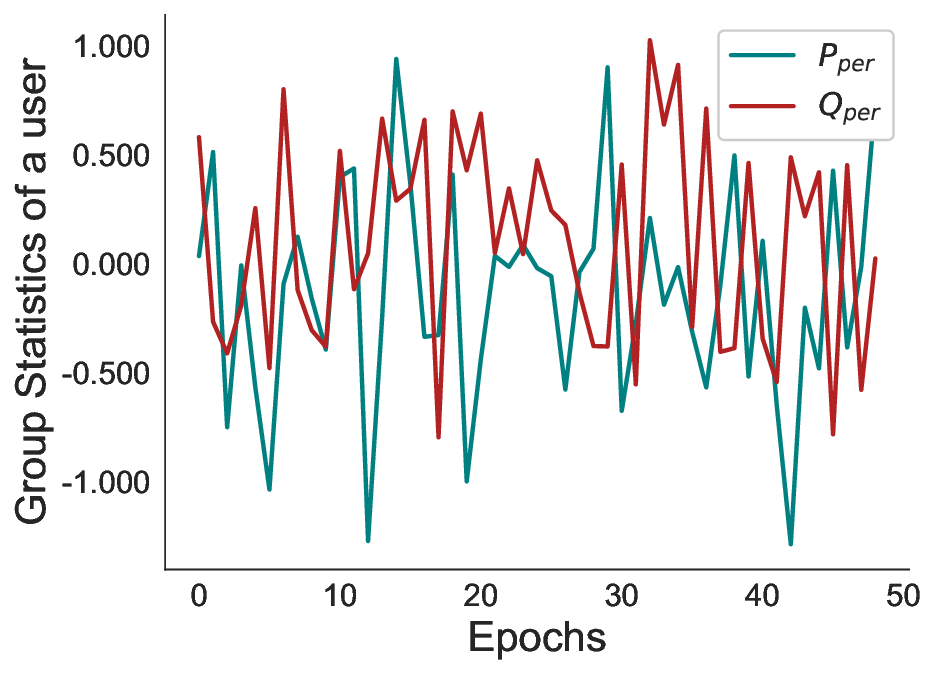}
\includegraphics[width=.23\textwidth]{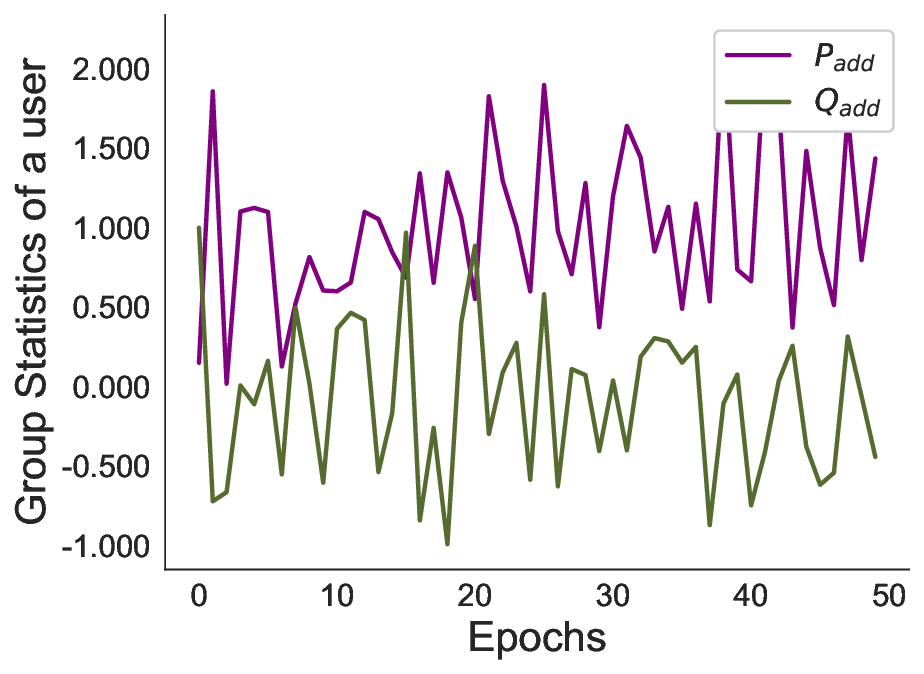}\quad
\caption{Effect of Local Differential Privacy (LDP) on group statistics for a given user of group $S_0$ for fixed $\beta$, $\delta$, $\lambda$ and $\sigma$ combination. \textbf{Left to Right:} ML-100K (G) ($P_{per}/Q_{per}$, $P_{add}/Q_{add}$), ML-1M (G) ($P_{per}/Q_{per}$, $P_{add}/Q_{add}$)}
\label{fig:GroupStatLDP2}
\end{figure*}

\section{Local Differential Privacy} \label{ldp}
In this section, we present the brief introduction of Local Differential Privacy (LDP) mechanism. LDP is a crucial technique employed to ensure the protection of sensitive information during collection and analysis \cite{Dwork14}. With the growing emphasis on safeguarding user privacy, LDP has gained significant attention. In a typical LDP scenario, a group of users shares private values with an untrusted third-party aggregator, who aims to obtain statistical insights into the distribution of these values among the users. LDP guarantees that the disclosure of individual user information is limited by applying a randomized algorithm, denoted as $\mathcal{A}$, to the private value $v$. The perturbed value $\mathcal{A}(v)$ is then shared with the aggregator for statistical inference. To satisfy $\epsilon$-local differential privacy, the randomized algorithm $\mathcal{A}\left(\cdot\right)$ must adhere to the following inequality for any two arbitrary input private values $v$ and $v'$:
\begin{equation}\label{eq:LDP_definition}
\operatorname{Pr}[\mathcal{A}(v)=y] \leq e^\epsilon \operatorname{Pr}\left[\mathcal{A}\left(v^{\prime}\right)=y\right]
\end{equation}
where $y \in range(\mathcal{A})$. Here, $\epsilon \geq 0$ is the privacy budget. Smaller $\epsilon$ indicates better privacy protection. \par
In F$^{2}$PGNN, local model gradient $\nabla \Theta _{u}$ is computed according to Eq.(\ref{eq:scale}), which may still reveal some private information about the user \cite{Zhu2019}. Hence, to enhance privacy, we employ LDP technique to  $\nabla \Theta _{u}$. We considered the randomized algorithm $\mathcal{A}$, applied to $v = \nabla \Theta _{u}$ as follows:
\begin{equation}\label{eq:clip_noise}
    \mathcal{A}(v) = clip\left(v, \delta\right)+ Laplace(0, \lambda)
\end{equation}
where $clip\left(v, \delta\right)$ limits the value of $v$ with the clipping threshold $\delta$. Here, $\lambda$ is the variance of the zero-mean Laplace noise and a larger value of $\lambda$ guarantees strong privacy protection. Upon completing the clip and randomization processes, inferring raw user behaviors from the gradients becomes considerably more challenging. Subsequently, the user client transmits the randomized local model gradient $\mathcal{A}(v)$ to the server.
\section{Experimental Settings}
\subsection{Dataset Summary}  \label{dataset_summary}
The number of users and items indicates the number of rows and columns in the rating matrix, and the number of ratings indicates the observed values in this matrix. The rating levels represent the range of ratings in different datasets. The sensitive attribute ``activity" of the users is defined using average rating threshold.
\begin{table}[!h]
\centering
\resizebox{\columnwidth}{!}{%
\begin{tabular}{cccccc}
 \hline
 \hline
 \multirow{2}*{Dataset} & \multirow{2}*{Users} & \multirow{2}*{Items} & \multirow{2}*{Ratings} & Ratings & Sensitive \\
 & & & & Levels & Feature \\
\hline
 
\multirow{2}*{ML-100K} & \multirow{2}*{943} & \multirow{2}*{1,682} & \multirow{2}*{1,00,000} & \multirow{2}*{1-5} & Gender \\
& & & & & Activity \\
\hline

\multirow{2}*{ML-1M} & \multirow{2}*{6,040} & \multirow{2}*{3,706} & \multirow{2}*{10,00,209} & \multirow{2}*{1-5}  & Gender \\
& & & & & Activity \\
\hline
Amazon & \multirow{2}*{5,515} & \multirow{2}*{13,509} &\multirow{2}*{4,84,141} & \multirow{2}*{1-5} &\multirow{2}*{Activity}\\
-Movies& & & & &  \\
\hline
\hline
\end{tabular}
}
\caption{Dataset's summary statistics.}\label{Table:datasummary}
\end{table}

\subsection{Why filtering n-core data?} \label{n-core}
 It might happen that the source of unfairness in recommendation may be the data imbalance in the respective groups. Hence, filtering $n$-core data ensures that while dealing with the fairness constraint, each user has at least $n$ samples. Using the 20-core data reduced the number of users from $943$, $6040$, $5515$ to $917$, $6022$, $5497$ whereas the number of items reduced from $1682$, $3706$, $13509$ to $937$, $3043$, $13457$ and the ratings reduced from $100000$, $1000209$, $484141$ to $94443$, $995154$, $458450$ for ML-100K, ML-1M and Amazon Movies respectively. \par
 
\subsection{Hyperparameter Settings} \label{Hyperparameter}
We utilize Graph Attention Networks (GAT) as the chosen GNN model for all our experiments. We use dot product to implement the rating predictor. The ratio of model dropout is fixed to $0.2$. The hidden dimension $\left ( h \right )$ for both user and item embeddings learned by GNN are set to $64$. For the baseline and F$^{2}$PGNN, we choose SGD as the optimization algorithm and its learning rates from the set $\left \{ 0.001, 0.01, 0.05 \right \}$. The batch dropout rate $\left ( K \right )$ is set to either $0.1$ or $0.5$. The total number of epochs varies from $20$ to $50$ as per the model convergence rate over different datasets. We consider the fairness budget $\beta$ in the set $\left \{ 0.3, 0.5, 0.7, 0.9 \right \}$. \par 
\section{Additional Experimental Results} 
\subsection{Privacy-Fairness tradeoff} \label{additional_ldp}

We investigate the effect of LDP on the fairness metric for gradient clipping threshold $\delta = 0.4$ and the strength of Laplacian noise $\lambda = 0.15$. Figure \ref{fig:LDPAnalysis2} illustrates the effect of LDP on fairness and privacy trade-off. It is evident that, F$^{2}$PGNN preserves the fairness effectively across all datasets, whereas, for Amazon Movies (A), at $\beta = 0,0.3$ the trend gets disturbed.

\subsection{Effect of LDP on $\mathcal{G}_{stat}$} \label{ldp_gstat}
Figure \ref{fig:GroupStatLDP2} shows the additional results on the remaining datasets (i.e ML-100K (G) and ML-1M (G)) to illustrate how the local group statistics $\left(\mathcal{G}_{stat}^{u}\right)$ for a particular user hide the users' true demographic features through LDP.

\section{Complexity Analysis}
As discussed in Section \ref{sec:PrivacyProtection}, each user needs to encrypt, at worst $m$ items in each communication round for private graph expansion, which can be done in $\mathcal{O}(m)$. Moreover, local group statistics updates require each user to scale the gradients by the scalar $L$, which depends on the number of demographic groups, i.e. $G$. This corresponds to the time complexity of $\mathcal{O}(G)$. Hence, the computational cost for each epoch at the central server is $\mathcal{O}(NmG)$ in addition to the communication head of \textit{FedAvg}, where $N =\left | U_K \right |$ is the number of participating users. In practice, $G$ is a very small integer as compared to other model parameters, which leads to smaller extra complexity.   
\end{document}